\documentclass[journal=jacsat,manuscript=article]{achemso}

\usepackage[version=3]{mhchem} 

\usepackage{xcolor}
\usepackage{braket}

\author{Kishan S. Menghrajani}
\altaffiliation{Current address: School of Physics and Astronomy, Monash University, Wellington Raid, Clayton, 3800, Victoria, Australia}
\email{kishansmresearch@gmail.com}
\author{Adarsh B. Vasista}
\affiliation[Exeter University]
{Department of Physics and Astronomy, Stocker Road, University of Exeter, Devon EX4 4QL, United Kingdom}
\altaffiliation{Current address: Department of Physics, Indian Institute of Science Education and Research, Bhopal 462066, India}
\author{Wai Jue Tan}
\affiliation[Exeter University]
{Department of Physics and Astronomy, Stocker Road, University of Exeter, Devon EX4 4QL, United Kingdom}
\author{Philip A. Thomas}
\affiliation[Exeter University]
{Department of Physics and Astronomy, Stocker Road, University of Exeter, Devon EX4 4QL, United Kingdom}
\author{Felipe Herrera}
\affiliation{Department of Physics, Universidad de Santiago de Chile, Av. Victor Jara 3493, Santiago, Chile}
\alsoaffiliation{Millennium Institute for Research in Optics, Concepci\'{o}n, Chile}
\author{William L. Barnes}
\affiliation[Exeter University]
{Department of Physics and Astronomy, Stocker Road, University of Exeter, Devon EX4 4QL, United Kingdom}
\email{w.l.barnes@exeter.ac.uk}

\title[finesse title]
  {Molecular Strong Coupling and Cavity Finesse}

\abbreviations{IR,NMR,UV}

\begin{document}


\begin{abstract}
Molecular strong coupling offers exciting prospects in physics, chemistry and materials science. Whilst attention has been focused on developing realistic models for the molecular systems, the important role played by the entire photonic mode structure of the optical cavities has been less explored. We show that the effectiveness of molecular strong coupling may be critically dependent on cavity finesse. Specifically we only see emission associated with a dispersive lower polariton for cavities with sufficient finesse. By developing an analytical model of cavity photoluminescence in a multimode structure we clarify the role of finite-finesse in polariton formation, and show that lowering the finesse reduces the extent of the mixing of light and matter in polariton states. We suggest that the detailed nature of the photonic modes supported by a cavity will be as important in developing a coherent framework for molecular strong coupling as the inclusion of realistic molecular models.
\end{abstract}


\textbf{TOC Graphic}
\begin{figure}[!h]
\includegraphics{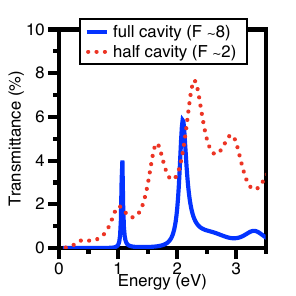}
\end{figure}


When molecules are placed inside an optical microcavity, the strong interaction between molecular resonances and cavity modes leads to the formation of hybrid states called polaritons - states that inherit characteristics of both the optical cavity modes and the molecules from which they are formed~\cite{Ebbesen_ACSAccounts_2016_49_2403,Herrera_JCP_2020_152_100902}.
This process, known as molecular strong coupling, has been extensively explored in the context of both excitonic~\cite{Lidzey_Nature_1998_395_53,Schwartz_PRL_2011_106_196405,Polak_ChemSci_2020_11_343} and vibrational resonances~\cite{Shalabney_NatComm_2015_6_5981,Long_ACSPhot_2014_2_130,Takele_PCCP_2021_23_16837}. In the context of vibrational strong coupling there is at present much excitement owing to the prospect of modifying chemical processes~\cite{Ebbesen_ACSAccounts_2016_49_2403,Yuen-Zhou_PNAS_2019_116_5214,Herrera_JCP_2020_152_100902,Garci-Vidal_Science_2021_373_eabd0336,Hirai_CPC_2020_85_1981,Thomas_Science_2019_363_615,Ahn_Science_2023_380_6650}, despite an incomplete understanding of the underlying science~\cite{Vurgaftman_JPCL_2020_11_3557,Imperatore_JCP_2021_154_191103,Vurgaftman_JCP_2022_156_034110}. A range of `cavity' structures have been explored, most frequently planar (Fabry-Perot) optical microcavities in which the molecules are located between two closely spaced metal or dielectric mirrors. Planar microcavities have dominated molecular strong coupling studies for many years, in both excitonic and vibrational regimes. However, such structures do not offer good access to the molecules involved, thereby limiting their applicability to cavity modified chemistry. Alternative `open' geometries have been explored, including surface plasmon modes~\cite{Baieva_JCP_2013_138_044707,Torma_RepProgPhys_2015_78_013901}, dielectric microspheres \cite{Vasista_NL_2020_20_1766}, and surface lattice resonances~\cite{Yadav_NL_2020_20_5043,Verdelli_JPCC_2022_126_7143}. More recently so-called `cavity-free' geometries have been investigated~\cite{Georgiou_JPCL_2020_11_9893,Thomas_JPCL_2021_12_6914,Canales_JCP_2021_154_024701,georgiou2023strong}, and extensive mode splitting observed. These cavities do not use metallic or dielectric multi-layer (DBR) mirrors, instead they rely on reflection from the interface of the molecular material with another dielectric to produce optical modes. Whilst some of the reports concerning open cavities have noted changes to the molecular absorption, it remains to be seen whether such structures can be used to control chemistry effectively. Since modification of photoluminescence is a more stringent measure of strong coupling than reflectance, transmittance, absorbance and scattering~\cite{Wersall_NL_2016_16_551,Wersall_ACSPhot_2019_6_2570,Vasista_Nanoscale_2021_13_14497,arXiv_2302.00023}, here we chose to explore the photoluminescence process for open, half and full cavities, in an attempt to gain better insight into cavity-free strong coupling. In doing so we identify an additional requirement that needs to be met for effective molecular strong coupling, one that highlights the vital role of cavity finesse.

\begin{figure*}
\includegraphics[width=1.0\columnwidth]{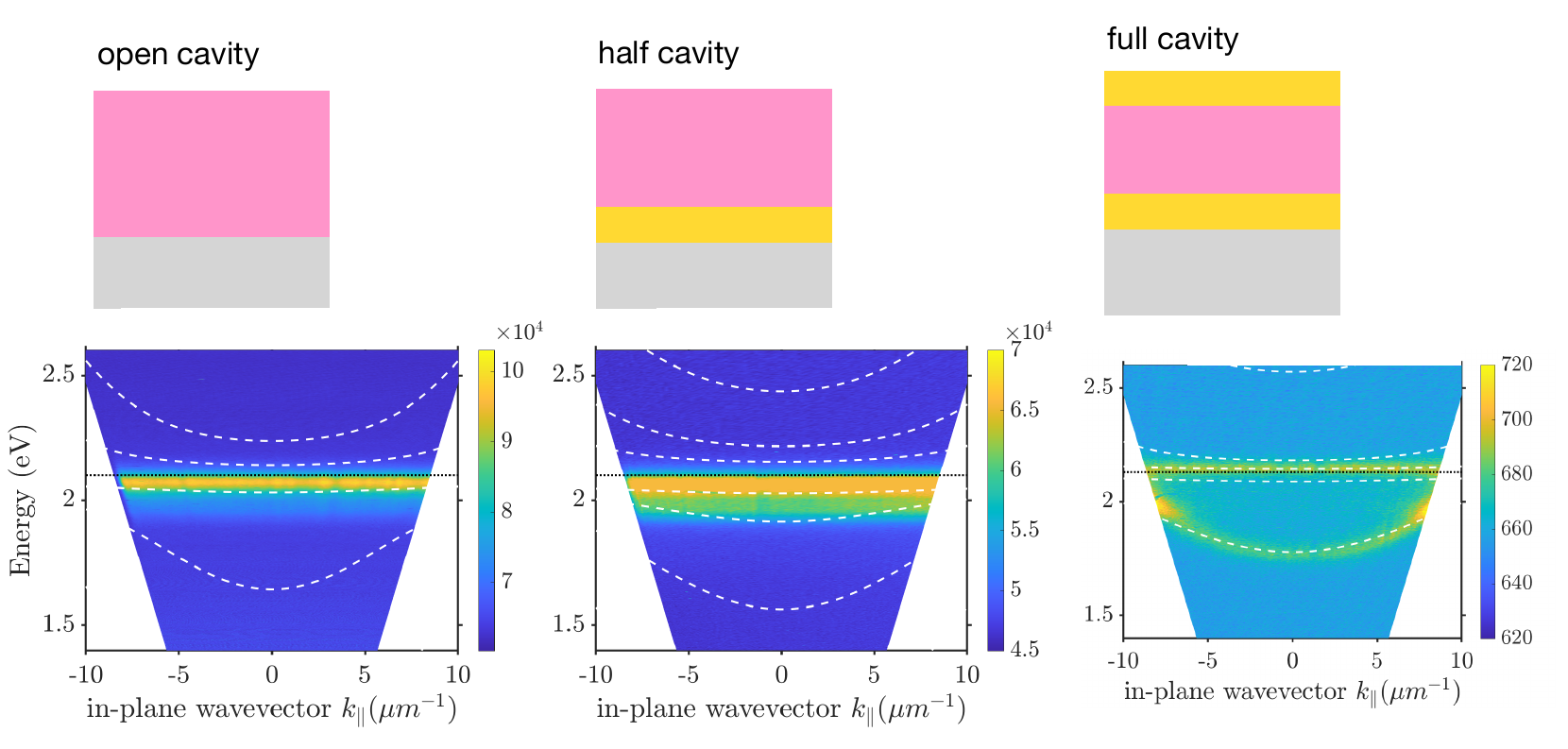}
\caption{\textbf{Schematic of Cavity Structures, and measured Photoluminescence dispersion:} \textbf{Left: Open Cavity} consisting of a TDBC-doped polymer (PVA) film on a silicon substrate, doped-PVA thickness $\sim$340 nm. \textbf{Centre: Half Cavity}, as (a) but now a thin gold film is included between the substrate and the dye-doped polymer, doped-PVA thickness $\sim$600 nm. \textbf{Right: Full Cavity}, similar to (b) but with the addition of a top gold layer to form the second mirror of a closed cavity, the cavity thickness is $\sim$400 nm. Photoluminescence spectra were acquired from each sample as a function of collection angle and the data are plotted here in the form of a dispersion diagram. The white dashed lines in each PL plot show the positions of the polaritons determined from the coupled oscillator model. Further data for these samples are shown in the SI. (Note that the (non-dispersive) emission in the vicinity of the molecular resonance is likely to be due to uncoupled molecules and weakly emissive dark states.)}
\label{fig:schematic of cavities and PL}
\end{figure*}

\vspace{0.3cm}
In the work reported here we made use of three different planar cavity structures, shown in the top row of figure \ref{fig:schematic of cavities and PL}: (left) an open cavity, i.e. a layer of polymer containing dye molecules supported by a silicon substrate; (centre) a half-cavity, similar to (a), but with the addition of a metallic (gold) mirror between the substrate and the dye-doped polymer, and; (right) a full-cavity, similar to the half-cavity but now with a second metallic mirror added to the top of the structure. For a more extended description of the optical modes supported by the different structures see Supplementary Information (SI) sections 4-6. We made use of the J-aggregated dye TDBC (5,5$^\prime$,6,6$^\prime$-tetrachloro-1,1$^\prime$-diethyl-3,3$^\prime$-di(4-sulfobutyl)-benzimidazolocarbocyanine), either dispersed in the polymer PVA, or deposited using a layer-by-layer approach~\cite{Vasista_Nanoscale_2021_13_14497}.
We used a silicon substrate, and made use of gold for the metallic mirrors; further details of fabrication are given in SI section 11. We measured photoluminescence and reflectance spectra as a function of polar angle, thereby enabling us to construct dispersion diagrams. We analysed our experimental data using a transfer matrix model to calculate the reflectance, transmittance and absorption, whilst we made use of a coupled oscillator model to determine the modes of each system; again, details of both are given in the SI (see sections 14 and 13 respectively). To enter the strong coupling regime the collective Rabi splitting, $\Omega_R$, should be greater than the mean of the cavity and molecular spectral widths, $K$ and $\Gamma$ respectively (see also the note in the SI, section 1), which we can write as~\cite{Rider_CP_2021_62_217},
\begin{equation}
\Omega_R > (K+\Gamma)/2.
\end{equation} 
\label{eq:usual_SC_condition}
\noindent However, as we will see below, satisfying this condition does not guarantee strong coupling as witnessed by photoluminescence; instead we find that we need to place another condition on the finesse of the cavity modes.

Photoluminescence (PL) and reflectance spectra were acquired as a function of wavelength and angle. For reflectance measurements a white-light source was coupled to an objective lens (100x, 0.8 NA) and focused on to the sample. The reflected light was then collected using the same objective lens and projected onto the Fourier plane~\cite{43}. For PL measurements, a 532 nm (green) diode-laser source was focused onto the sample and the PL was collected by the same objective lens in the back-scattering configuration. Details of the optical setup are provided in section 3 of the SI.

For each system we determined the extent of the anti-crossing (collective Rabi-splitting), $\Omega_R$, and the width (FWHM) of the cavity mode, $K$. The cavity mode-width was estimated from the calculated reflectance of `no-resonance' systems, where `no-resonance' means that the oscillator strength was set to zero, see for example figure S2, panel (c), and see section 7 of the SI. To estimate the collective Rabi splitting we adopted an iterative process as follows. A coupled oscillator model was used to reproduce the dispersion of the polariton modes, based on the `no-resonance' mode positions and an estimate for the Rabi frequency. The results from the coupled oscillator model were then plotted on top of the experimental (reflection and PL) data, and on top of the calculated reflection data. The value of the Rabi frequency was then adjusted to give a simultaneous best `by eye' match to the different data sets (see also section 13 of the SI). The reflectance data are thus key to determining the Rabi splitting since they exhibit both UP and LP features, however, we also ensured that the polariton positions we predict are consistent with the PL data. The full data sets are shown in figures S2-S8, a subset of the (reflectance) data are shown in figure \ref{fig:Refl_Rabi} below.

\begin{figure*}
\includegraphics[width=1.0\columnwidth]{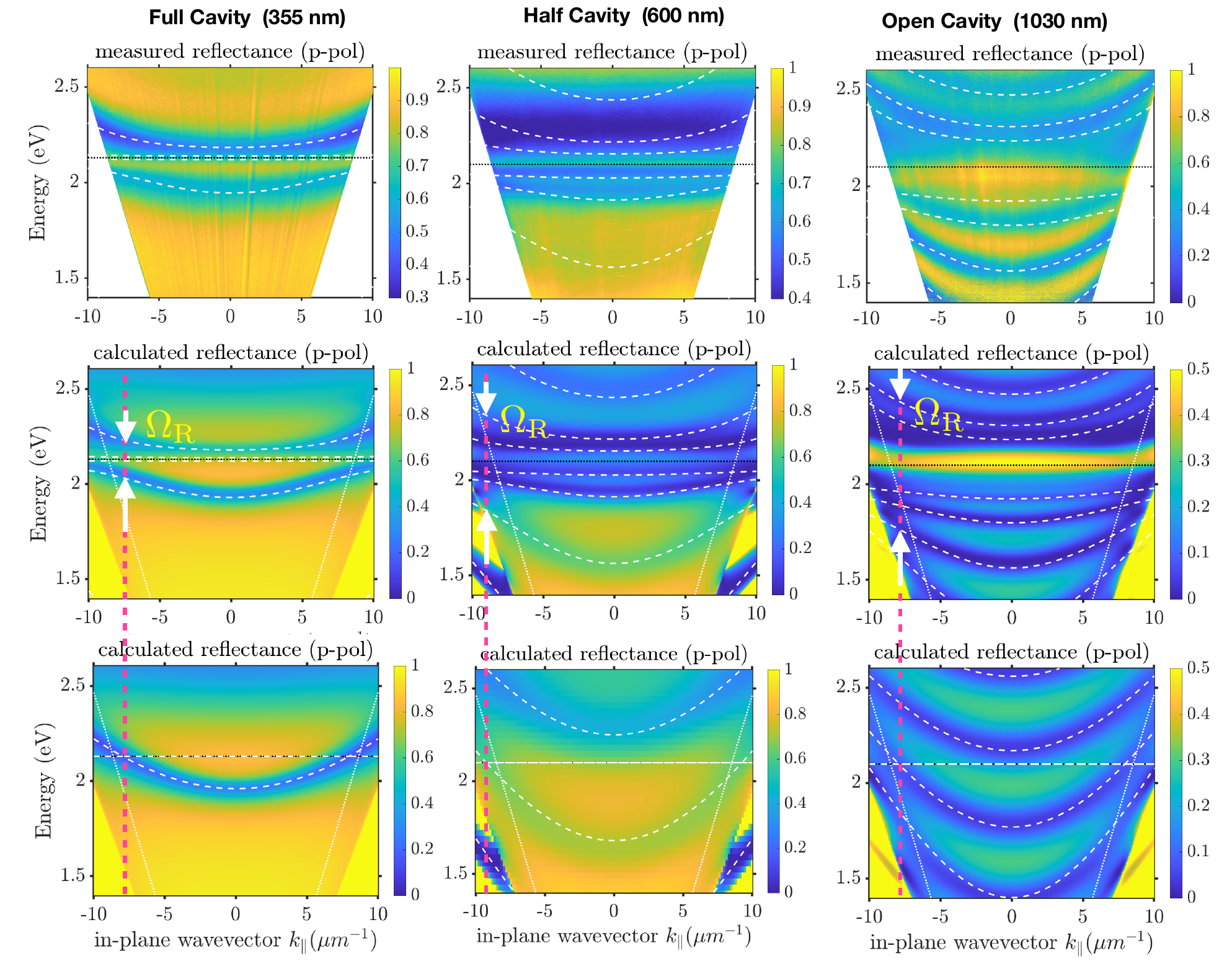}
\caption{\textbf{Estimating the Rabi splitting:} In each column the top panel shows the measured reflectance, the middle column the calculated reluctance, and the lower panel the calculated reflectance with the oscillator strength set to zero. In addition, we have superimposed on each panel the results from the best `by eye' match of a coupled oscillator model to the data. The left-hand column is for a full cavity (355 nm thick), the middle column for a half-cavity (600 nm thick), the right-hand column for an open cavity (1030 nm thick). The crossing point of the key cavity mode with the excitonic resonance is shown as a magenta dotted line in each column. In each of the middle panels the arrows indicate the extent of the Rabi-splitting. Further information is given in figures S2-S8.}
\label{fig:Refl_Rabi}
\end{figure*}

Let us now return to consider figure 1 in more detail. In the second row of figure \ref{fig:schematic of cavities and PL}, we show examples of the collected photoluminescence from the three types of cavity in the form of dispersion diagrams.
Here the PL spectra have been plotted as a function of frequency and in-plane wavevector $k_{\parallel} = \frac{2\pi}{\lambda} \sin\theta$ (where $\lambda$ is the wavelength of light and $\theta$ is the angle of incidence; the plane is that of the sample). Also shown as white dashed lines are the energies of the hybrid modes determined via the coupled oscillator model, further details are given in the SI. For each configuration a clear LP is present in reflectivity between 1.6 and 1.8 eV, but only in the case of the full cavity is PL associated with the LP seen. As we will show below, PL is only seen to be associated with the dispersive LP if the cavity finesse is sufficiently high.


Let us return now to discuss the PL data for each cavity configuration in turn. In the second row, left column of figure \ref{fig:schematic of cavities and PL} we show the measured photoluminescence dispersion. We observe a strong peak at 2.07 eV and a weaker peak at 1.97 eV. For reference a PL spectrum from a very thin (20 nm) TDBC film on Si is also shown in Supplementary Figure S9. The reference spectrum is very similar to that of the open cavity: for the open cavity case, the 2.07 eV peak is slightly broader and the 1.97 eV shoulder slightly stronger. The photoluminescence of the open cavity is also non-dispersive. There is thus little if any sign that strong coupling has influenced the dye photoluminescence of this open cavity. One might argue that when the lower polariton mode at $k_{\parallel}=0$ is so far in energy from the un-modified photoluminescence that no change would be expected. However, we have observed elsewhere that this need not be the case~\cite{Vasista_Nanoscale_2021_13_14497}, provided there are phonon/vibrational modes that can scatter emission via the polariton.
Note that there is no dispersion of the PL in the vicinity of the dispersion curve where anti-crossing might be anticipated.
We repeated this experiment for a thicker ($\sim$1030 nm) TDBC film (Supplementary Figure S3) and once again found that the PL is only marginally modified (if at all) in the open cavity configuration.
To investigate the absorption in the TDBC layer we made further use of transfer matrix modelling, the results are shown in figure S2 of the SI. Although there is a significant change in the absorption (in the bulk this would be a single peak) it is clear that the distortion is not due to the presence of polaritons. Instead the doublet feature in absorption in this seemingly simple sample arises from the complex interplay between absorption and the impedance that the TDBC-layer presents to incoming light~\cite{Tan_JCP_2021_154_024704}.
This is consistent with our previous modelling of the absorption of cavity-free strong coupling with a broad spectrum dye~\cite{Thomas_JPCL_2021_12_6914}, which showed modified absorption but no clearly resolvable polariton modes.


The half cavity case (centre column of figure \ref{fig:schematic of cavities and PL}) appears very similar to the open cavity case (a larger PL peak at 2.07 eV and a smaller PL peak at 1.97 eV), but with a smaller difference between the magnitudes of the two peaks. Again, there is no clear mapping of the PL onto the position of the polariton modes. Calculated data for the absorption in the TDBC layer are shown in the SI, figure S4 panel (e). There is again a significant change in the absorption compared to that of the bulk, and further, compared to the case of the open cavity, there is now some indication of the absorption tracking the lower polariton mode, at least to some limited extent. Data collected from a 1630 nm thick half cavity sample (supplementary figure S5) also show a somewhat modified PL spectrum.


The measured photoluminescence dispersion for the full cavity (right hand column of figure 1) is significantly different from the open and half cavity cases, the PL clearly tracking the lower polariton mode. Calculated data for the absorption in the TDBC are shown in SI figure S8, panel (e). As for the PL, there is now a very significant change in the absorption that also clearly tracks the polariton modes.

\begin{figure}[htb!]
\centering
\includegraphics[width=0.9\linewidth]{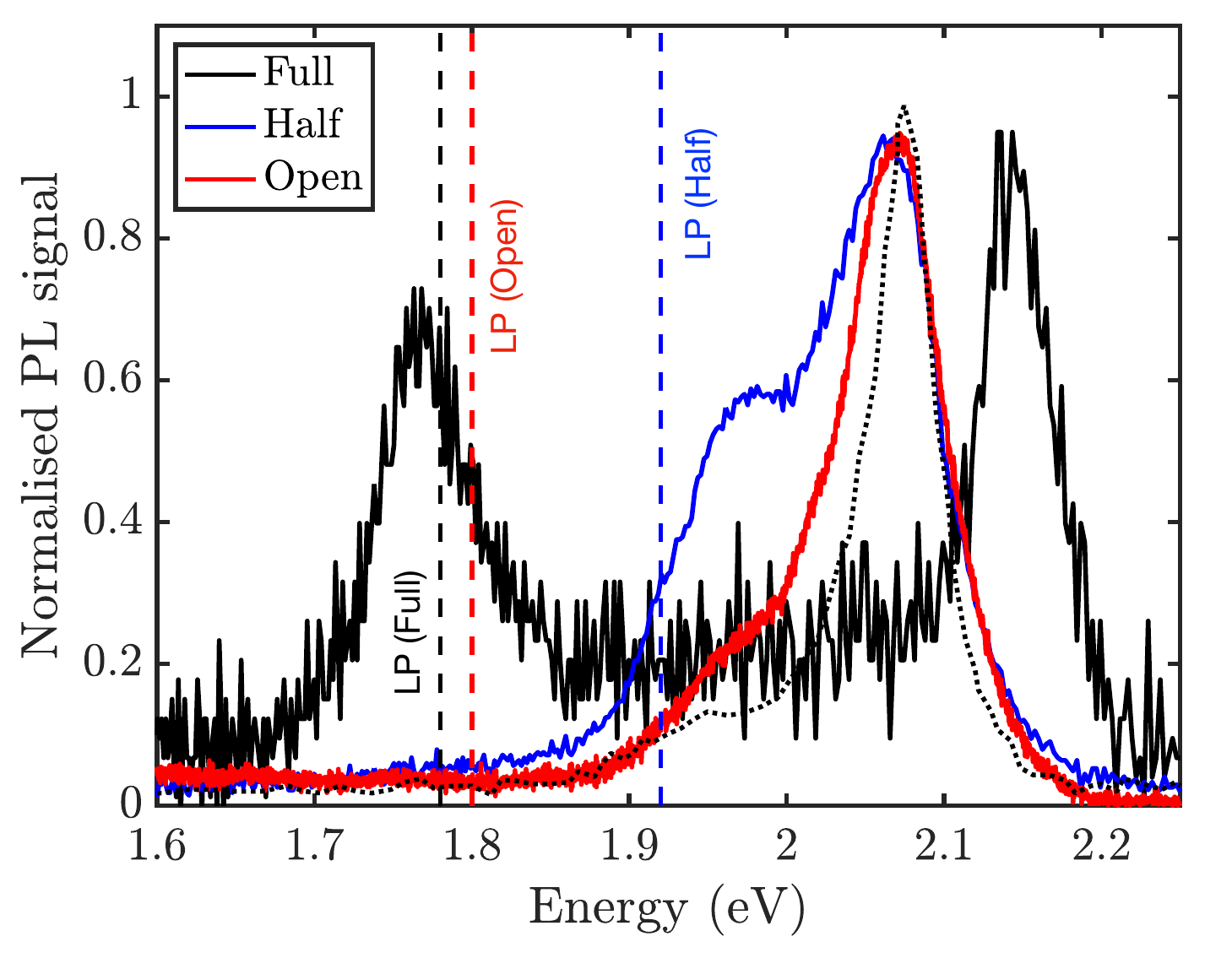}
\caption{\textbf{PL line spectra}.
PL for normal emission for an open cavity (1030 nm sample, red line), a half cavity (600 nm sample, blue line), and full cavity (355 nm sample, black line). The PL data have been normalised and scaled to lie between values of 0 and 1.
The position of the lower polariton modes for $k_{\parallel}=0$ are shown as vertical dashed lines, the lower polaritons shown here are associated with the mode that crosses the TDBC absorption energy in the zero-oscillator strength dispersion, see panels (c) in supplementary figures  S3, S4, and S8. The thin film reference PL data set (dotted line) was acquired from a thin film of TDBC on a silicon substrate, see supp info section S7. Note the difference in the peak position of the `bare' PL for the full cavity data when compared to the open and half cavity data. Part of this difference can be attributed to the fact that the TDBC for this full cavity was made using the layer-by-layer technique (see supp info), whilst for the other two data sets the TDBC was on the polymer host PVA, again, see supp info section S7.}
\label{fig:PL_comparison}
\end{figure}

\vspace{0.2cm}
It is useful at this point to compare the line spectra for the PL, for which we have chosen $k_{\parallel}=0$.  Figure \ref{fig:PL_comparison} shows the PL spectra for another set of open, half and open cavities with different thicknesses than those in fig. \ref{fig:schematic of cavities and PL}, but with equivalent spectral behaviour (full dispersion data in supplementary figures S3, S4, and S8). We have indicated the position of the lower polariton of the least detuned cavity mode at $k_{\parallel}=0$. As a reference `uncoupled' case, a thin film of TDBC (20 nm) on Si is used, which has a strong PL peak at 2.07 eV with a weak shoulder around 1.96 eV.

For the {\it open cavity}, the 1.97 eV PL shoulder is clearer than in the thin film and the 2.07 eV peak is broader and the PL remains non-dispersive. For the {\it half cavity}, the 1.97 eV PL peak is substantially enhanced compared to the open cavity, but there is still no clear PL dispersion. The thicker half-cavity PL is slightly dispersive. The changes at 1.97 eV might be weakly linked to the (lower) polariton modes supported by this structure, see figure S4 (d,f). The absorption spectrum shows signs of being modified.

In contrast, for the {\it full cavity} both  PL and absorption clearly track the lower polariton. We also note that, as commonly found~\cite{Bellessa_PRL_2004_93_036404,Agranovich_PRB_2003_67_085311,Schwartz_CPC_2013_14_125}, PL is observed from polaritons at energies lower than the molecular resonance energy, i.e. we do not see any PL associated with the upper polariton. Looking at the dispersion of the PL from the lower polariton we see that it is not uniform: PL is typically produced by the relaxation of reservoir states through the loss of vibrational energy~\cite{Coles_JPCA_2010_114_11920} so that PL emission is strongest when the difference between the bare molecular resonance energy (reservoir) and the polariton branch are equal to the energy of a vibrational mode~\cite{Vasista_Nanoscale_2021_13_14497}.

\vspace{0.3cm}
We have compared the photoluminescence from dye molecules (TDBC aggregates) located in three different cavity configurations: open, half and full cavities. In all three cases the reflectivity data indicate the strong coupling regime has been reached, see the fifth column of Table 1. The calculated absorption shows a somewhat different picture, with changes in the absorption by the TDBC for all three cavity types, but only in the case of the full cavity does the absorption track the (lower) polariton fully.
We also observe changes in PL for all three samples, but again it is only for the full cavity that the PL maps onto the lower polariton. 
The behaviour we observe in PL, figure \ref{fig:PL_comparison}, is reminiscent of the transition from weak to strong coupling observed in reflection measurements~\cite{Thomas_NanoLett_2020}, where an initially uncoupled peak broadens before splitting into two clearly resolvable peaks.
In PL the upper polariton is not observed due to non-radiative relaxation, so instead a lower polariton branch eventually becomes distinct from the uncoupled PL peak.
Ordinarily, the transition from weak to strong coupling is observed by increasing the number of molecules in a cavity, for example by using photochromic molecules~\cite{Schwartz_PRL_2011_106_196405,Thomas_NanoLett_2020}. Here, however, similar behaviour has instead been observed by modifying the cavity structure. This leads to a number of questions: how can we quantify the change in these structures that has caused this transition into the strong coupling regime, and what is the threshold for the observation of strong coupling in PL?

\vspace{0.3cm}
Previously we have looked at whether absorption by the dye is modified in the strong coupling regime, and -- as here -- we found that for an open cavity there was some modification~\cite{Thomas_JPCL_2021_12_6914}. However, that study made use of a broad spectrum dye (whereas TDBC is narrow-band -- with a spectral width of $\Gamma$ = 0.07 eV) -- and no PL measurements were undertaken. Other work looking at strong coupling between dye molecules and particle plasmon modes found that there was clear PL arising from the lower polariton in the strong coupling regime~\cite{Wersall_ACSPhot_2019_6_2570}.

\begin{table*}[htb!]
\centering
\begin{tabular}{|p{1.1cm}| p{1.8cm}|p{1.8cm}| p{2.1cm} | p{2.3cm} | p{1.9cm} | p{1.8cm} |}
\hline
\hline
Cavity\newline(nm)&$\Delta\omega\,$(eV)\newline{(FSR)}&$K\,$(eV)\newline{(mode-width)} & $\Omega_R\,$(eV)\newline{(Rabi splitting)}& $2\Omega_R/(K+\Gamma)$\newline{($>$ 1 for SC )}&$Q$\newline{(Q-factor)}& $\mathcal{F}$\newline{\textbf{(Finesse)}}\\
\hline
\hline
\textbf{Open}& & & & & & \\
\hline
340 &1.20$\pm0.02$&0.50$\pm0.02$&0.50$\pm0.03$&1.75$\pm0.12$&4.2$\pm0.2$&\textbf{2.4$\pm0.2$}\\
\hline
1030 &0.38$\pm0.01$&0.17$\pm0.01$&0.67$\pm0.04$&5.58$\pm0.57$&12.4$\pm0.7$&\textbf{2.2$\pm0.2$}\\
\hline
\hline
\textbf{Half}&&&&&&\\
\hline
600 &0.64$\pm0.01$&0.25$\pm0.01$&0.50$\pm0.03$&3.13$\pm0.27$&8.4$\pm0.4$&\textbf{2.6$\pm0.1$}\\
\hline
1630 &0.25$\pm0.01$&0.10$\pm0.01$&0.35$\pm0.02$&4.12$\pm0.40$&21$\pm2$&\textbf{2.5$\pm0.4$}\\
\hline
\hline
\textbf{Full}&&&&&&\\
\hline
330 &0.73$\pm0.01$&0.13$\pm0.01$&0.16$\pm0.01$&1.60$\pm0.19$&14$\pm4$&\textbf{5.6$\pm0.6$}\\
\hline
355 &0.75$\pm0.01$&0.11$\pm0.01$&0.20$\pm0.01$&2.22$\pm0.27$&19$\pm2$&\textbf{6.2$\pm0.6$}\\
\hline
400 &0.76$\pm0.01$&0.09$\pm0.01$&0.27$\pm0.02$&3.38$\pm0.49$&23$\pm2$&\textbf{8.4$\pm1.2$}\\
\hline
\hline
\end{tabular}

\caption{\textbf{Spectral parameters for the different cavities}. The relevant figures are as follows: open cavity, figures S2 and S3; half cavities, figures S4 and S5; full cavities, S6 - S8.}
\label{tab:spectral_parameters}
\end{table*}

\vspace{0.3cm}
What are we to make of our results? To address this question we looked for a correlation between our findings and the cavity parameters, e.g. $Q$-factor. In table~\ref{tab:spectral_parameters} we have brought together the parameters for our different structures. As noted in the introduction, a conventional criterion for strong coupling is that $2\Omega_R>(\Gamma + K)$. Based on the data in table \ref{tab:spectral_parameters} and the fact that for TDBC, the spectral width is $\Gamma$ = 0.07 eV, all of our samples meet this criterion (see fifth column of table~\ref{tab:spectral_parameters}). It thus appears that this criterion is not the full story.  

We next focus our attention on a previously ignored parameter, the cavity finesse, $\mathcal{F}$, given by $\mathcal{F}=\Delta\omega/K$, where $\Delta\omega$ is the free spectral range (FSR), i.e. the frequency separation of adjacent modes, see also SI section 7. Looking now at the penultimate column in table~\ref{tab:spectral_parameters}, there appears to be a clear correlation between the behaviour we see in our PL results and the cavity finesse. We find that for low finesse structures, $\mathcal{F}\sim 2$, the PL does not track the (reflectance) lower polariton, even when the Rabi splitting exceeds the linewidths. For higher finesse values, $\mathcal{F}> 4$, the PL does track the (reflectance) lower polariton.

Based on our analysis of the cavity finesse (see table \ref{tab:spectral_parameters}) we suggest an additional criterion, to supplement the usual strong coupling criterion, (1), we suggest,

\begin{equation}
2\Omega_R > (K+\Gamma),\, \textrm{and},\, \mathcal{F}>\sim 4. 
\end{equation} 
\label{eq:new_condition}

\begin{figure}[t]
\centering
\includegraphics[width=0.9\linewidth]{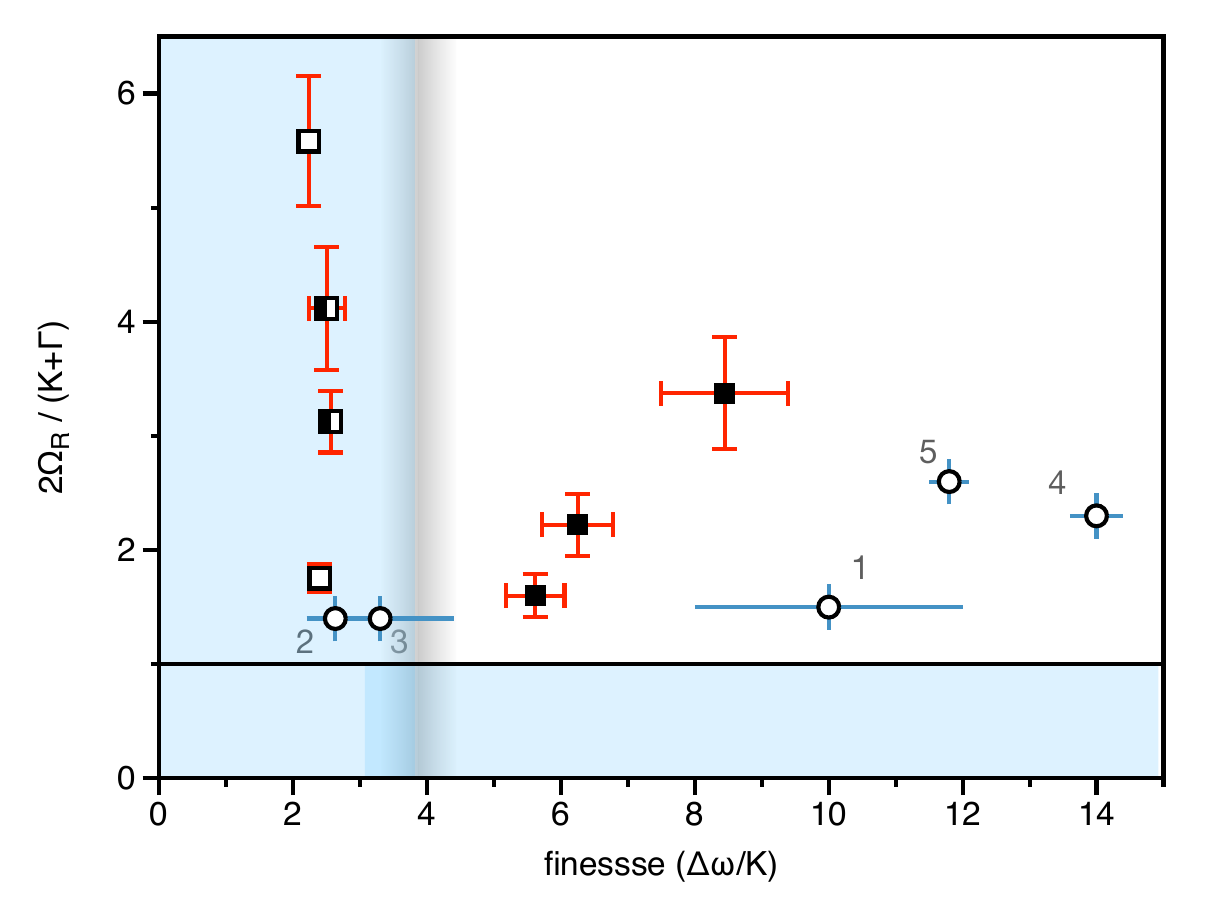}
\caption{\textbf{Strong value of coupling parameter space}.
Values of $2\Omega_R/(K+\Gamma)$ and $\Delta \omega /K$ (finesse) for each of the cavities we investigated. The error bars (red) are derived from the data in table \ref{tab:spectral_parameters}. Data from open cavities are indicated by an open square, from half cavities by a half-filled square, and from full cavities by a filled square.
Also shown are data points (together with associated error bars, in this case blue) based on other reports, for details see main text.
The horizontal line at $2\Omega_R /(K+\Gamma) = 1$ indicates the `usual' strong coupling condition. The vertical blurred line indicates our suggested criterion based on cavity finesse. Our 2D criterion is satisfied in the region of white background.}
\label{fig:KRW}
\end{figure}

\noindent We have plotted our data in this form in figure \ref{fig:KRW}; solid and grey lines indicate the two criteria. For the data associated with the cavities that we have investigated here (data points with red error bars), we see that the open and half cavities are all such that the free spectral range is too low. In contrast, the values for the three full cavities have both sufficient finesse \textbf{\textit{and}} sufficient coupling strength for effective strong coupling.\\

To explore these ideas further we also plot in figure \ref{fig:KRW} data extracted from a number of reports in the literature. Point 1 is for the dye-coated plasmonic nano-prism reported by Wersall \textit{et al.}~\cite{Wersall_NL_2016_16_551}. It is clear that for the plasmonic particle system investigated by Wersall \textit{et al.} effective strong coupling was achieved, and their PL data confirm this. Point 2 is for the open dielectric cavity of Thomas \textit{et al.}~\cite{Thomas_JPCL_2021_12_6914}. As for the open cavities we have explored here, it is perhaps marginal whether this system has attained the effective strong coupling regime. In this case, coupling to a higher-order electromagnetic mode would greatly lower $K$, potentially pushing this system into the effective strong coupling regime.
Point 3 is for the dielectric microsphere of Vasista \textit{et al.}~\cite{Vasista_NL_2020_20_1766}. It is clear in this case that whilst the coupling strength is adequate, the finesse is too low. It may be that a reduction in microsphere size (resulting in an increase in $\Delta\omega$) would allow the effective strong coupling regime to be reached. 
Points 4 and 5 are planar Fabry-Perot cavities used in two studies that report modifications to chemical reactions due to vibrational strong coupling. Point 4 corresponds to the work of Thomas \textit{et al.}~\cite{Thomas_Science_2019_363_615}, whilst point 5 corresponds to the work of Ahn \textit{et al.}~\cite{Ahn_Science_2023_380_6650}. In both cases the effective strong coupling regime is comfortably reached. More information on these data is given in the SI.

\begin{figure*}[t]
\centering
\includegraphics[width= 0.9\textwidth]{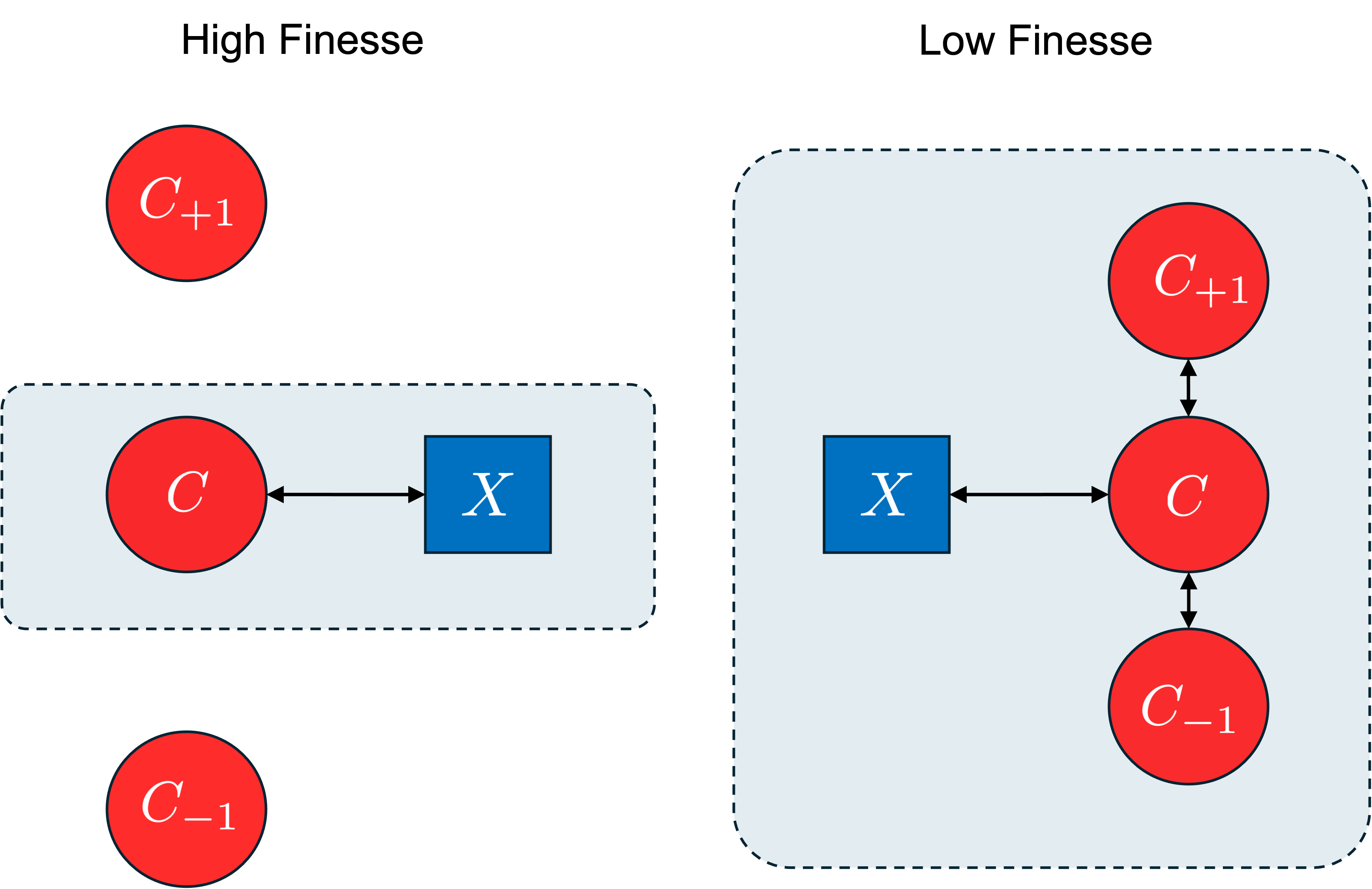}
\caption{\textbf{Schematic of effect of multiple photonic modes}.
The schematic shows three photonic modes $\{C_{-1},C,C_{+1}\}$ coupling to a single molecular resonance. For high finesse cavities (left) the dominant interaction is between the molecular exciton mode ($X$) and the energetically-closest photonic mode ($C$). In low-finesse cavities (right), the exciton content is shared between neighbouring partially overlapping cavity modes, thus modifying the character of polariton emission signals.}
\label{fig:schematic}
\end{figure*}

How might we understand this requirement of a lower limit on the finesse in multi-mode cavities for effective strong coupling? Molecular strong coupling relies on the coherent exchange of energy between a set of molecular resonators and a cavity mode. If the finesse is too low then the molecular resonators can interact with multiple cavity modes simultaneously, thus changing the properties of the lower and upper polariton states around the molecular resonance. In figure \ref{fig:schematic}  we schematically compare the type of mode mixing present in high and low finesse cavities. Three photonic modes $\{C_{-1},C,C_{+1}\}$ are shown in order of increasing energy, interacting with an excitonic mode ($X$) that is resonant with the central mode $C$.  For high finesse, only the $C$ and $X$ modes exchange energy, giving the usual single-mode picture of light-matter interaction. In low finesse cavities, the exciton content is shared among multiple partially overlapping cavity modes, which changes the emission properties of the lower polariton in the central spectral region.

The consequences of this multi-mode mixing process in cavity PL can be understood by breaking down the cavity emission into a multi-step process in which molecular dipoles are first pumped incoherently from the ground state $S_0$ to the lowest excited state $S_1$, via UV excitation $S_0\rightarrow S_n$ followed by ultrafast radiationless relaxation $S_n\rightarrow S_1$ \cite{Muccini_2000}, they then give up their excitation energy to the vacuum cavity field creating individual confined photons, which finally decay to the far field through the mirrors at rate $K$, thereby generating the PL signal.

To model PL in a multi-mode cavity with tunable finesse, we extend the theoretical analysis of Herrera and Spano \cite{Herrera_PRA_2017_95_053867,Herrera2017-prl}, by explicitly modelling the probability of  exciting molecular dipoles pumped incoherently at rate $W$ and including multiple cavity modes at discrete frequencies $\omega_q$ ($q$ an integer). Assuming that the coupled light-matter is such that no coherence between polaritonic eigenstates is present and depletion of the ground state due to incoherent pumping is negligible, the stationary PL spectrum is then given by,
\begin{equation}\label{eq:PL spectrum}
S_{PL}(\omega)= \pi W \,\sum_j \frac{ X_j^T}{NW X_j^T+(NW+K/2)C_j^T}\,C_j^q\,L_j(\omega),
\end{equation}
where the discrete index $j$ labels each polaritonic eigenstate in the single excitation manifold (including dark states~\cite{Herrera_PRA_2017_95_053867}),  $X_j^T$ is the total exciton content of the $j$-th eigenstate , $C_j^q$ is the photon content of the $j$-th state in the $q$-th cavity mode and $C_j^T=\sum_q C_j^q$ is the total photon content summed over all cavity modes. $L_j(\omega)$ is a normalized Lorentzian response function with central eigenfrequencies $\omega_j$ and bandwidth $\Gamma_j$ (FWHM), which for simplicity we set to $\Gamma_j=K$ for all states. $N$ is the number of molecular dipoles. The derivation of Eq. (\ref{eq:PL spectrum}) is given in Section 10 of the SI.

\begin{figure*}[t]
\centering
\includegraphics[width=\linewidth]{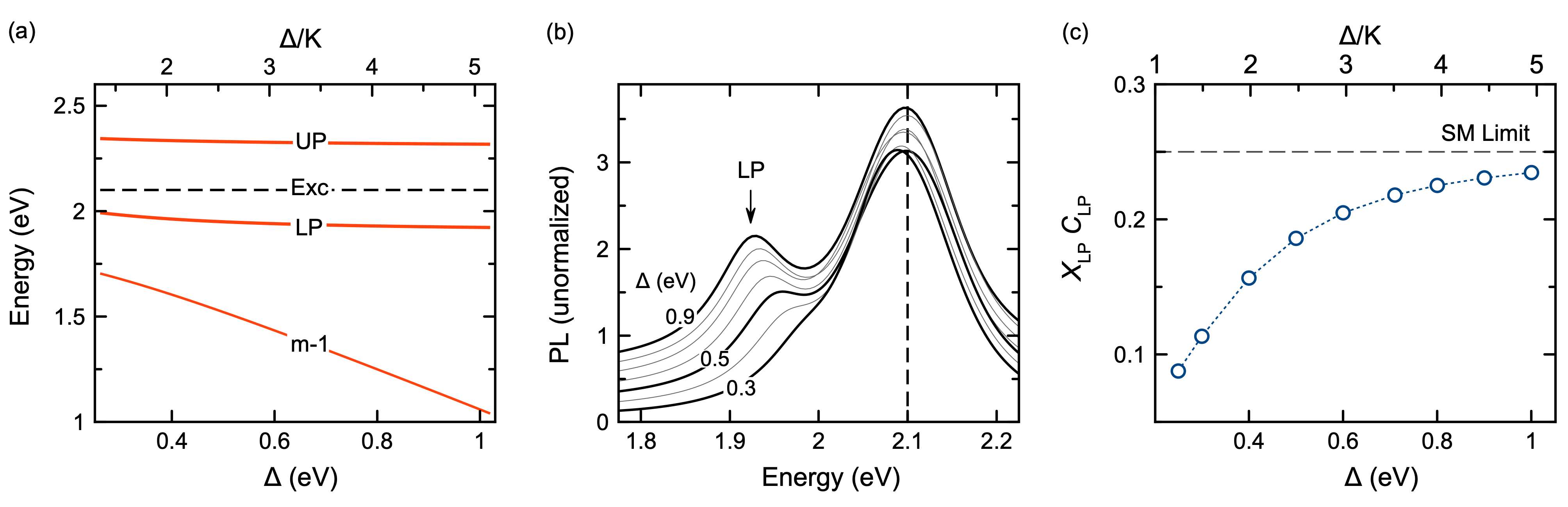}
\caption{\textbf{PL in a two-mode cavity}.
(a) Calculated polariton spectrum of a two-mode cavity as a function of the mode energy separation $\Delta$. The central $q=m$ mode strongly couples to the molecular resonance at $\omega_e=2.1$ eV and is kept fixed in energy. The lower $q=m-1$ mode approaches the molecular resonance with decreasing $\Delta$. (b) Simulated PL spectra of the two-mode cavity in panel (a), for an ensemble of $N$ in-homogeneously broadened molecular dipoles centered at 2.1 eV (vertical dashed line, $\sigma=0.021$ eV). Curves are labelled by the value of $\Delta$. LP denotes lower polariton. (c) Calculated PL spectral weight $X_{\rm LP}C_{\rm LP}$ at the LP energy as a function of $\Delta$, for the same parameters in panel (b). The single-mode limit $X_{\rm LP}C_{\rm LP}=1/4$ is marked with a dashed horizontal line. (Note that the spectral weight product $X_{\rm LP}\,C_{\rm LP}$ comes from Eq. (\ref{eq:PL spectrum}) where here we have set $j={\rm LP}$, $q=m$, and we have dropped the other indices for notational simplicity.) In all cases we set $N=150$, $\sqrt{N}g=0.2$ eV, $K=0.2$ eV, and $NW/K=10^{-4}$. Curves are averaged over 650 disorder configurations. In panels (a) and (c) the finesse $(\Delta/K)$ is plotted on the upper abcissa.}
\label{fig:PL multimode}
\end{figure*}

At this point it is worth looking at the structure of Eq. \eqref{eq:PL spectrum}, specifically the physical significance of the factors involved. The quantity we are calculating is the strength of the PL signal, $S_{PL}$. First, for low pumping rates (no saturation) the strength of the PL signal is directly proportional to the pumping rate $W$. Second, the factor $X_j^T/[NW X_j^T+(NW+K/2)C_j^T]$ gives the probability that the lower polariton state is occupied after incoherent pumping of the dipoles. Third, the factor $C_j^q$ gives the photon content of the lower polariton state, it is this `fraction' of the state that is available to yield photons that leak out of the cavity sample. Finally, as noted above, $L_j(\omega)$ is a (Lorentzian) lineshape function.

Figure \ref{fig:PL multimode}(a) shows the calculated emission spectrum of an idealized two-mode cavity with tunable finesse. The central mode $q=m$ is kept at exact resonance with a molecular transition at $2.1$ eV, and the energy separation $\Delta$ to the lower $q=m-1$ mode is varied. The positions of the LP, UP and exciton lines are marked. An ensemble of $N$ molecular dipoles is equally coupled to both cavity modes with local coupling strength $g=\Omega_R/\sqrt{N}=0.2/\sqrt{N}$ (eV), which gives a Rabi spliting of $\Omega_R=0.4$ eV for large $\Delta$. We choose this coupling magnitude such that the usual single-mode picture of strong coupling holds for the resonant mode ($K=0.2$ eV, $\Gamma=0.05$ eV).

Figure \ref{fig:PL multimode}(b) shows the  PL signal at the LP energy, calculated using Eq. (\ref{eq:PL spectrum}), for different mode separation energies $\Delta$,  ($\Delta$ is proportional to finesse). A Gaussian distribution of molecular transition frequencies is assumed ($\bar\omega_e$ = 2.1 eV, $\sigma = 21$ meV,  FWHM = 0.05 eV). As the finesse changes, the central emission feature near the bare molecular resonance remains largely unaltered, in qualitative agreement with the experimental comparison in figure \ref{fig:PL_comparison}, however the PL from the lower polariton changes substantially. The  lower polariton feature in PL emerges with increasing inter-mode separation $\Delta$, as controlled by the spectral weight product $X_{\rm LP}\,C_{\rm LP}$ in Eq. (\ref{eq:PL spectrum}) (where we have set $j={\rm LP}$, $q=m$, and we drop the other indices for notational simplicity). This product can be understood as the degree of admixing between light and matter at the LP energy. When a light-matter eigenstate at frequency $\omega_j$ is either purely photonic ($X_j^T=0$) or purely excitonic ($C_j^q=0$), the PL signal is strongly suppressed. 

For a single-mode resonant cavity with a spectrally homogeneous ensemble of molecules ($\sigma=0$), the lower polariton state in strong coupling has $X_{\rm LP}\approx 1/2$ and $C_{\rm LP}\approx 1/2$, which sets the optimal mixing limit for the PL spectral weight $C_{\rm LP}X_{\rm LP}\approx 1/4$. Figure \ref{fig:PL multimode}(c) shows how the PL spectral weight of the LP state monotonically decreases from this asymptotic upper limit as the energy separation between modes decreases. In addition to the change in spectral weight, for intermode energy separation $\Delta$ comparable or smaller than the collective Rabi coupling $\sqrt{N}g$, we also expect level pushing of the LP from below towards the molecular resonance, for constant  light-matter coupling strength. The combination of reduced spectral weight and level pushing gives the spectral progression shown in figure \ref{fig:PL multimode}(b).

In summary, we investigated the photoluminescence, reflectance, and absorption associated with a range of dye-doped cavities. Whilst in all the configurations we studied we saw evidence of strong coupling in the reflectance data, this was not true for PL. We only saw PL associated with a dispersive LP for the full cavities. For the other cavity configurations that we examined we saw non-dispersive emission in the vicinity of the excitonic resonance. Whilst a full analysis of the details of this non-dispersive emission is beyond the scope of the present study, it is likely to include emission from uncoupled molecules (aggregates) and weakly emissive dark states. We compared a variety of spectral parameters with the behaviour we observed in photoluminescence, and found a correlation with cavity finesse. We developed a theoretical model of photoluminescence under strong coupling so as to specifically include coupling between adjacent photonic modes, coupling that arises when the photonic modes are of low finesse. Our model provides a conceptually straightforward explanation of our results; the dominant effect of this coupling in low finesse situations is that coupling between adjacent photonic modes reduces the matter content of the polariton, thus reducing the probability of polariton emission. The absence of PL associated with the dispersive LP in low finesse situations is thus a result of the reduced (excitonic) content of the LP due to `sharing’ of the exciton content among more than one cavity mode.

Whilst our aim was to better understand strong coupling in open cavities, we have arrived at a more general conclusion about effective strong coupling: that in addition to the usual condition on the coupling strength, an additional condition based on the cavity finesse needs to be met. This is perhaps not so surprising. The `traditional' strong coupling criterion is based on considering a single molecular resonance interacting with single cavity mode. A natural consequence is that the hybrid polariton modes that arise are half matter (exciton), half light (cavity mode). When more than one cavity mode is involved things can become more complicated and this simple 50/50 light-matter distribution no longer applies~\cite{Bhuyan_AOM_2023_12_2301383}. As we have shown here, when the finesse is low then neighbouring cavity modes can interact, leading to a reduction in the exciton content of a given polariton mode, see figure \ref{fig:PL multimode}. We suggest that in low finesse situations the `traditional' strong coupling criterion no longer ensures sufficient mode-mixing for all processes, e.g. photoluminescence, to be tied to the polariton modes. We expect the influence of finesse on strong coupling to apply to other optical microcavities, plasmonic nano-cavities etc., as well as infrared resonators. Future experiments using novel molecular cavity designs, as well as realistic microscopic quantum theory that includes the entire cavity mode profile, molecular dephasing and collective relaxation effects, will further refine our fundamental understanding of molecular strong coupling.

\begin{acknowledgement}

The authors acknowledge useful discussions with Marie Rider and William Wardley. K.S.M. acknowledges financial support from the Leverhulme Trust research grant ``Synthetic biological control of quantum optics''. K.S.M. also acknowledges the support of Royal Society International Exchange grant (119893R).
KSM, AV, PAT and WLB acknowledge the support of European Research Council through the photmat project 
(ERC-2016-AdG-742222 :www.photmat.eu). FH acknowledges the support of the Royal Society through the award of a Royal Society Wolfson Visiting Fellowship, and through grants Fondecyt Regular 1221420 and Millennium Science Initiative Program ICN17\_012. For the purpose of open access, the authors have applied a Creative Commons Attribution (CC BY) licence to any Author Accepted Manuscript version arising.
Data associated with these results can be found at https://doi.org/10.24378/exe.5306.
\end{acknowledgement}

\begin{suppinfo}

Strong coupling criteria; Optical modes of different cavity structures; Optical Fourier set-up; Open Cavities; Half Cavities; Full Cavities; Estimating the Free Spectral Range and the cavity mode linewidth; Photoluminescence from different thin TDBC layers, Literature Values; Microcavity PL with finite finesse; Sample preparation; Reflectance measurements; coupled oscillator models; transfer matrix modelling; Technical details re: TOC figure.

\end{suppinfo}

\bibliography{finesse}

\end{document}


\tableofcontents

\section{\textbf{Strong coupling criterion}}

A variety of criteria for the strong coupling regime are available in the literature~\cite{Rider_CP_2021_62_217}. The criterion is often expressed in words as the coupling rate exceeding the damping (molecular and cavity) rates. Whilst a formal discussion of the different criteria has some merit, in the present context the exact choice of which of the different criteria to use is not a key issue. What we attempt to do in the present work is to make a comparison between the criterion we chose ($\Omega_R = 2g_N>(K+\Gamma)/2$) and the emergence of PL associated with the lower polariton in our experiments.

\clearpage

\section{\textbf{Optical Modes of different cavity structures}}
It is worth clarifying what we mean here by `open', `half', and `full' cavities. It is perhaps easiest to begin with the full cavity. The \textbf{full cavity} structure has a high reflectivity metallic mirror both above and below the molecular layer and is thus capable of supporting optical modes that are largely confined within the molecular layer (there will be a small degree of leakage of the fields through the metal mirrors). For the \textbf{half cavity} there is one high reflectivity (metallic) mirror under the molecular layer, the upper mirror is simply the low reflectivity interface between the molecular layer and air. Such structures may support well-confined optical modes when the in-plane wavevector of the mode is greater than the wavevector in air, i.e. in the regime of total internal reflection. For optical modes of lower wavevector the interface with the air provides only a weak reflection so that the optical field is no longer as well confined, such modes are often referred to as half-leaky modes. For the \textbf{open cavity} employed here we have a different situation. The lower interface between the molecular layer and the silicon substrate provides a modest degree of reflectivity whilst the interface between the molecular layer and air provides a somewhat lower reflectivity. Modes within the light line (in-plane wavevector less than that in air) are poorly confined to the molecular material, such modes are often referred to as leaky modes. For modes with an in-plane wavevector greater than that in air total internal reflection yields much more tightly confined modes.

For both the half and open cavities the situation can be rather complex when more than one mode is supported, such that one or more modes lie inside the light-line (and are thus not well-confined) and one or more lie beyond the light-line~\cite{Menghrajani_ACSPhot_2020_7_2448}. 

\clearpage 

\section{\textbf{Optical Fourier set-up}}
Figure S1 shows a schematic of the experimental setup. To measure angle resolved reflectance, a white light source was focused onto the sample using a 0.8 NA 100x objective lens and the reflected signal was collected using the same lens. Lens L4 and L5, together, project the back-focal plane of the objective lens onto either the spectrometer or the camera, depending on the position of the flip mirror FM2. Lens L6 was a flip-lens used to project the real plane to the spectrometer/camera. For Photoluminescence (PL) measurements, a beam-expanded diode-pumped 532 nm source was focused onto the sample using the objective lens, and the PL was collected in the back-scattering configuration. The laser line was then rejected using spectral edge filters placed after lens L5. For k$\sim$0 excitation (normal incidence), we place a lens, L7, in the input path such that the laser was focused onto the back aperture of the objective lens, resulting in an approximate parallel beam with k$\sim$0 at the sample plane.

\begin{figure}[h!]
\centering
\includegraphics[width=\linewidth]{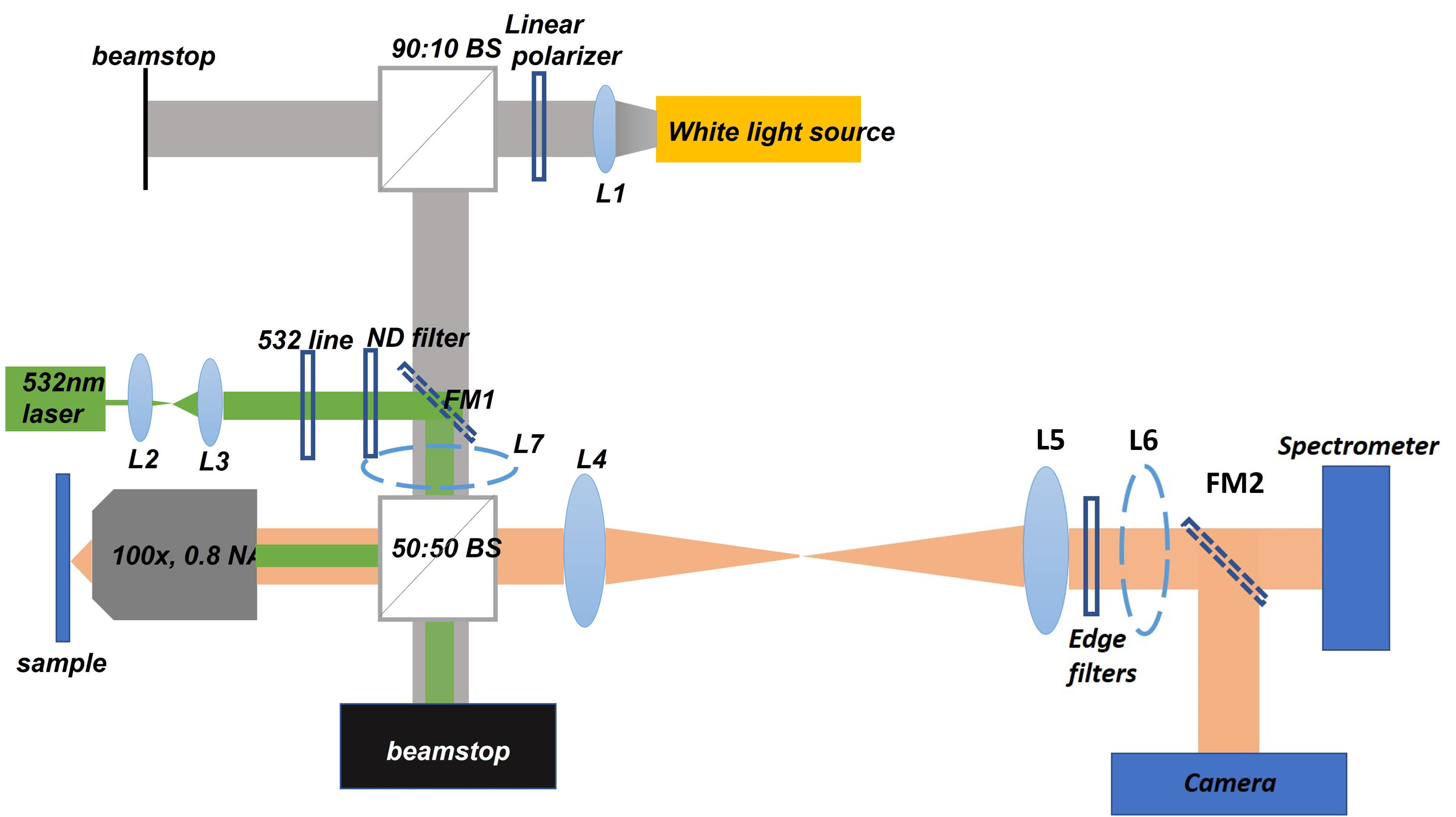}
\caption{Schematic of the Fourier optical setup}
\label{fig:SI_Fourier_setup}
\end{figure}

\clearpage

\section{\textbf{Open Cavities}}

\begin{figure}[htb!]
\centering
\includegraphics[width=1.00\linewidth]{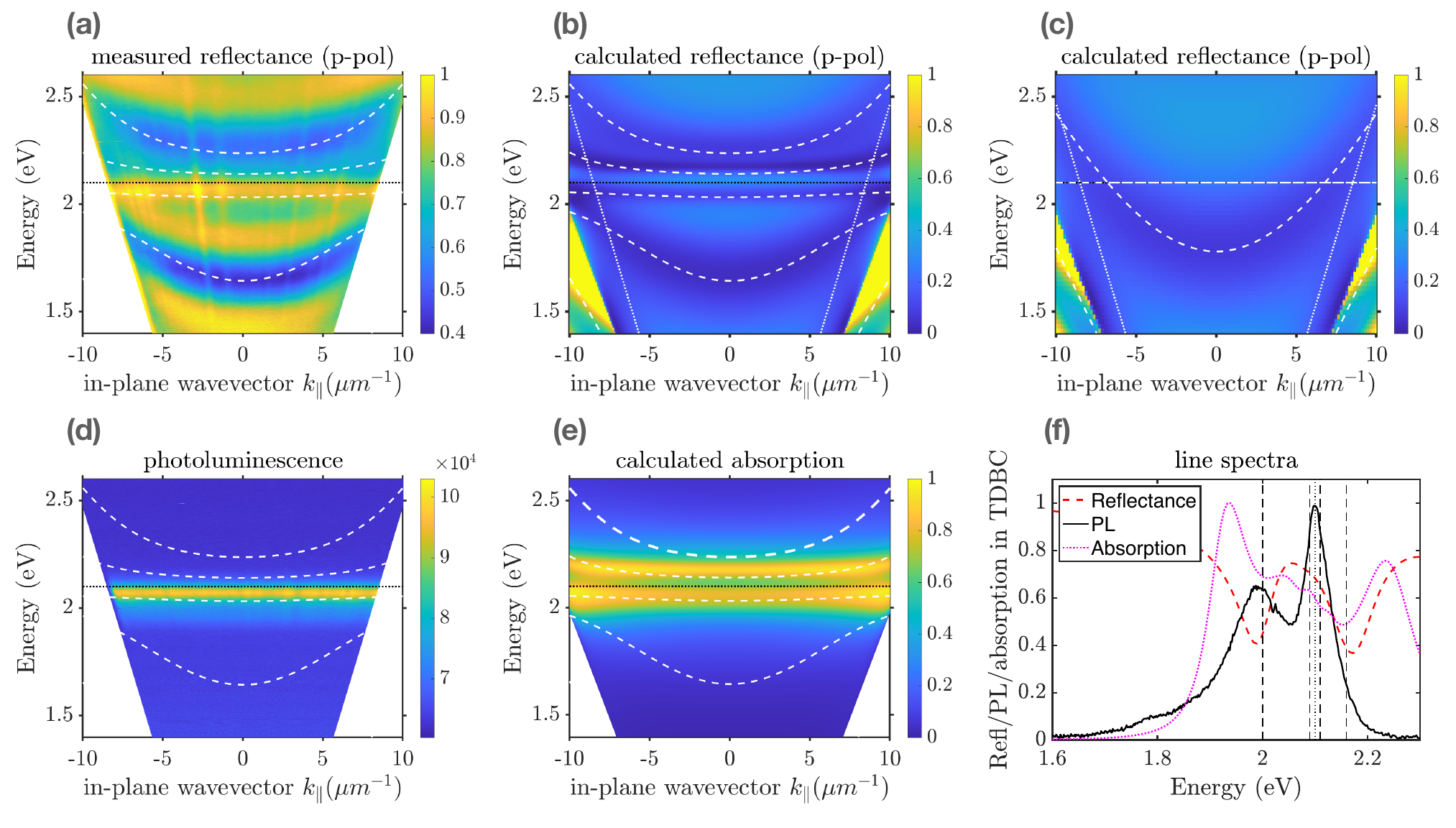}
\caption{\textbf{Open cavity}, polymer thickness $\sim$340 nm. (a) Experimentally measured dispersion, based on p-polarised reflectance, indicated as a colour plot (see methods re: normalisation).
(b) Calculated reflectance, the white dotted lines represent the limit of the collection numerical aperture used in the experiment (max angle = 55$^{\circ}$). (c) As panel (b) except that in the calculation we have set the oscillator strength of the dye to zero.
(d) Experimentally measured dispersion plot acquired from photoluminescence data. (e) Calculated absorption in the TDBC-doped layer. In panels (a-e) the polariton dispersion is shown, as determined from the coupled oscillator model described in the methods section. The polariton positions are shown as white dashed lines, and the molecular resonance (2.10 eV) by a black dotted line; just one cavity mode was used in the model (see panel (c)). (f) Normal incidence line spectra for calculated reflectance, measured PL and calculated absorption. The polariton positions (normal incidence) are indicated by black dashed lines.}
\label{fig:open_data_340nm}
\end{figure}

\vspace{0.3cm}
In figure \ref{fig:open_data_340nm} we show, in the upper row, dispersion data based on reflectance for the open cavity, i.e. for a $\sim$340 nm TDBC-doped PVA layer on a silicon substrate. On the left (a) we show the measured reflectance, in the middle the calculated reflectance (b), and in (c) we again show the calculated reflectance, but this time we set the oscillator strength of the dye to zero. In the lower row  we again show a dispersion plot, panel (d), this time based on the measured photoluminescence. In panel (e) we show the calculated absorption in the TDBC-doped PVA layer, whilst in panel (f) we show normal incidence line spectra for the calculated reflectance, the measured photoluminescence and the calculated absorption in the TDBC. The white dashed lines in panels (a-e) indicate the positions of the polariton modes as determined from our coupled oscillator model (see methods), the polariton positions at normal incidence are indicated in panel (f) by black dashed lines. Further details are given in the figure caption.

\vspace{0.3cm}
Looking first at the reflectance, panels (a) and (b), the presence of modes of the open cavity system is indicated by regions of low reflectivity (blue). Also shown are the positions of the polaritons as determined by our coupled oscillator model (details in methods). We see that overall there is a reasonable match between the experimental data, the calculated data, and the coupled oscillator model (which here employed a single cavity resonance). The calculated dispersion for a `no resonance' cavity and the `no resonance' coupled oscillator model is shown in panel (c). Here `no resonance' cavity and `no resonance' coupled oscillator model refer to the oscillator strength being set to zero.

From the modelling we determine that in this case there are three photonic modes to consider, with de-tunings (relative to the molecular resonance) of -1.5, -0.32, and +0.85 eV, having beta values of 1.7, 1.6 and 1.9 respectively ($\beta$ is a parameter that ensures the no-resonance cavity dispersion is correctly included in the coupled oscillator model). Note that these modes were determined by looking at data spanning a greater frequency range than shown here. The reduced oscillator strength is 0.14, and the splitting is $\Omega_R = 0.50\pm0.03$ eV. From figure~S2(c) we find the mode width to be $K=0.50\pm0.02$ eV. 

\begin{figure}[h!]
\centering
\includegraphics[width=1.00\linewidth]{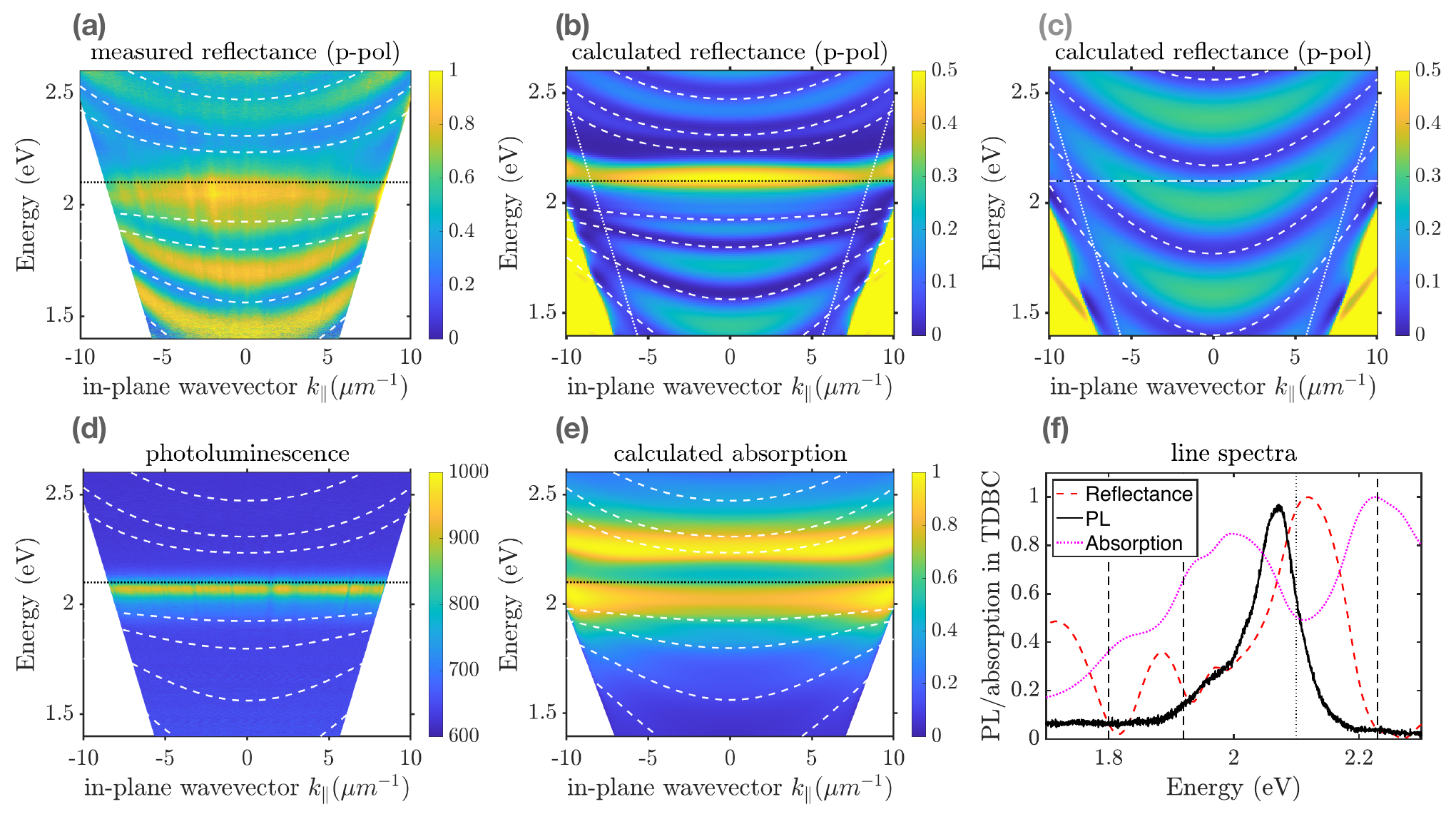}
\caption{{\textbf{Open Cavity (1030 nm)}.
In panel (a) we show an experimentally measured dispersion, based on p-polarised reflectance, indicated as a colour plot (see methods re: normalisation).
In panel (b) we show the calculated reflectance, the white dotted lines represent the limit of the collection numerical aperture used in the experiment. Panel (c) is the same as panel (b) except that in the calculation we have set the oscillator strength of the dye to zero.
In panel (d) we show an experimentally measured dispersion plot acquired from photoluminescence data. In panel (e) we show the calculated absorption in the TDBC-doped layer. In panels (a-e) the polariton dispersion is shown, as determined from the coupled oscillator model described in the methods section. The polariton positions are shown as white dashed lines, and the molecular resonance (2.10 eV) by a black dotted line; three cavity modes were used in the model (see panel (c)), and the polymer thickness was $\sim$1030 nm.} In panel (f) the polariton positions (normal incidence) are indicated by black dashed lines.}
\label{fig:open_data_1030}
\end{figure}

\vspace{0.3cm}
In figure \ref{fig:open_data_1030} we show, in the upper row, dispersion data based on reflectance for the open cavity, i.e. for a $\sim$1030 nm TDBC-doped PVA layer on a silicon substrate. The sample was similar to that used for the open cavity data shown in figure 2 of the main manuscript, the primary difference being that here the thickness of the cavity was 1030 nm rather than 340 nm used for the sample discussed in the main manuscript. On the left (a) we show the measured reflectance, in the middle the calculated reflectance (b), and in (c) we again show the calculated reflectance, but this time set the oscillator strength of the dye to zero. In the lower row  we again show a dispersion plot, panel (d), this time based on the measured photoluminescence. In panel (e) we show the calculated absorption in the TDBC-doped PVA layer, whilst in panel (f) we show normal incidence line spectra for the measured reflectance, the measured photoluminescence and the calculated absorption in the TDBC. The white dashed lines in panels (a-e) indicate the positions of the polariton modes as determined from our coupled oscillator model (see methods), the polariton positions at normal incidence are indicated in panel (f) by black dashed lines. Further details are given in the figure caption.

\vspace{0.3cm}
Looking first at the reflectance, panels (a) and (b), the presence of modes of the open cavity system is indicated by regions of low reflectivity (blue). Also shown are the positions of the polaritons as determined by our coupled oscillator model (details in methods). We see that overall there is a reasonable match between the experimental data, the calculated data, and the coupled oscillator model (which here employed three cavity resonances). The calculated dispersion for a `bare' cavity and the `bare' coupled oscillator model is shown in panel (c). Here `bare' refers to the oscillator strength in the Lorentz oscillator model being set to zero.\\ 

From the modelling we determine that in this case there four photonic modes to consider, with de-tunings of -0.7, -0.33, +0.07, and +0.46 eV, having beta values of 1.4, 1.2, 1.1 and 1.1 respectively. We determine that in this case the reduced oscillator strength is 0.25, the splitting is $\Omega = 0.67\pm0.04$ eV. 
Panel (d) shows the measured photoluminescence dispersion, we see that there is no sign of any influence of polaritons on the photoluminescence for this open cavity. Finally, to investigate the absorption we made further use of transfer matrix modelling, the results are shown in panel (e).

\clearpage

\section{\textbf{Half Cavities}}

\begin{figure}[htb!]
\centering
\includegraphics[width=1.00\linewidth]{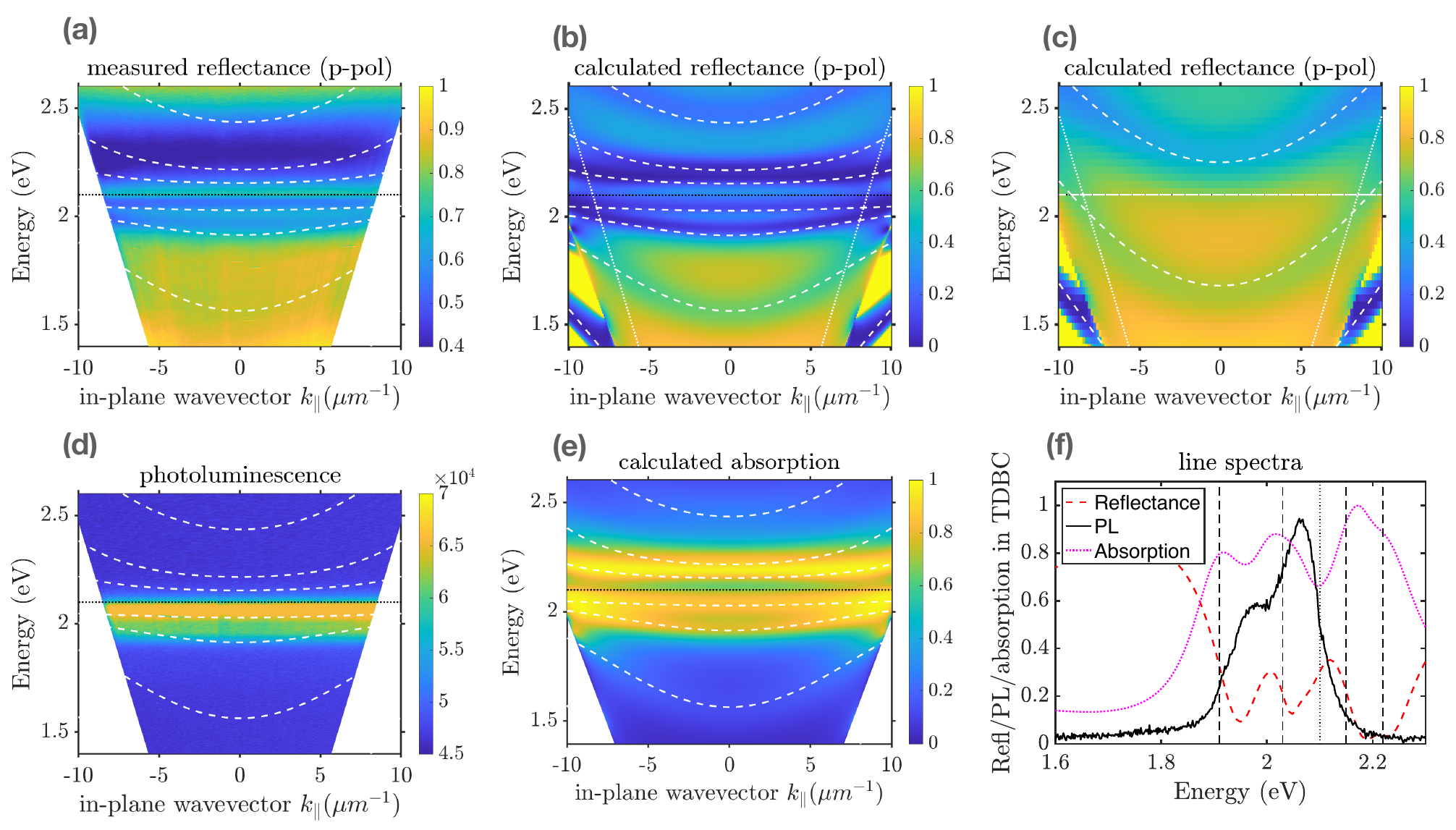}
\caption{\textbf{Half cavity}, polymer thickness $\sim600$ nm. (a) Experimentally measured dispersion, based on p-polarised reflectance, indicated as a colour plot (see methods re: normalisation). (b)  calculated reflectance, the white dotted lines represent the limit of the collection numerical aperture used in the experiment.  (c) As panel (b) except that in the calculation we have set the oscillator strength of the dye to zero.
(d) Experimentally measured dispersion plot acquired from photoluminescence data. (e) Calculated absorption in the TDBC-doped layer. In panels (a-e) the polariton dispersion is shown, as determined from the coupled oscillator model described in the methods section. The polariton positions are shown as white dashed lines, and the molecular resonance (2.10 eV) by a black dotted line; just one cavity mode was used in the model (see panel (c)). (f) Normal incidence line spectra for the calculated reflectance, the measured photoluminescence and the calculated absorption in the TDBC. The polariton positions (normal incidence) are indicated by black dashed lines.}
\label{fig:half_data_600}
\end{figure}

\vspace{0.3cm}
Figure \ref{fig:half_data_600} shows data from a 600 nm TDBC-doped PVA layer on a gold-coated silicon substrate. Looking first at the reflectance, panels (a) and (b), the presence of the modes of the half cavity system are again indicated by regions of low reflectivity (blue). Also shown are the positions of the polaritons as determined by our coupled oscillator model, where we made use of two cavity modes. We note that overall the match between the experimental data, the calculated data, and the coupled oscillator model are not as good as they were for the open cavity. The calculated dispersion for a `no-resonance' cavity and the `no-resonance' coupled oscillator model is shown in panel (c).\\

From the modelling we determine that in this case there are four photonic modes to consider, with de-tunings (relative to the molecular resonance of -1.1, -0.42, +0.15, and +0.80 eV, having beta values of 1.1, 1.1, 1.2 and 1.2 respectively. We determine that in this case the reduced oscillator strength is 0.14, and the splitting is $\Omega = 0.50\pm0.03$ eV. The width of the `no-resonance' cavity modes in this case is $K=0.25\pm0.01$ eV. 
Panel (d) shows the measured photoluminescence dispersion.

\begin{figure}[h!]
\centering
\includegraphics[width=1.00\linewidth]{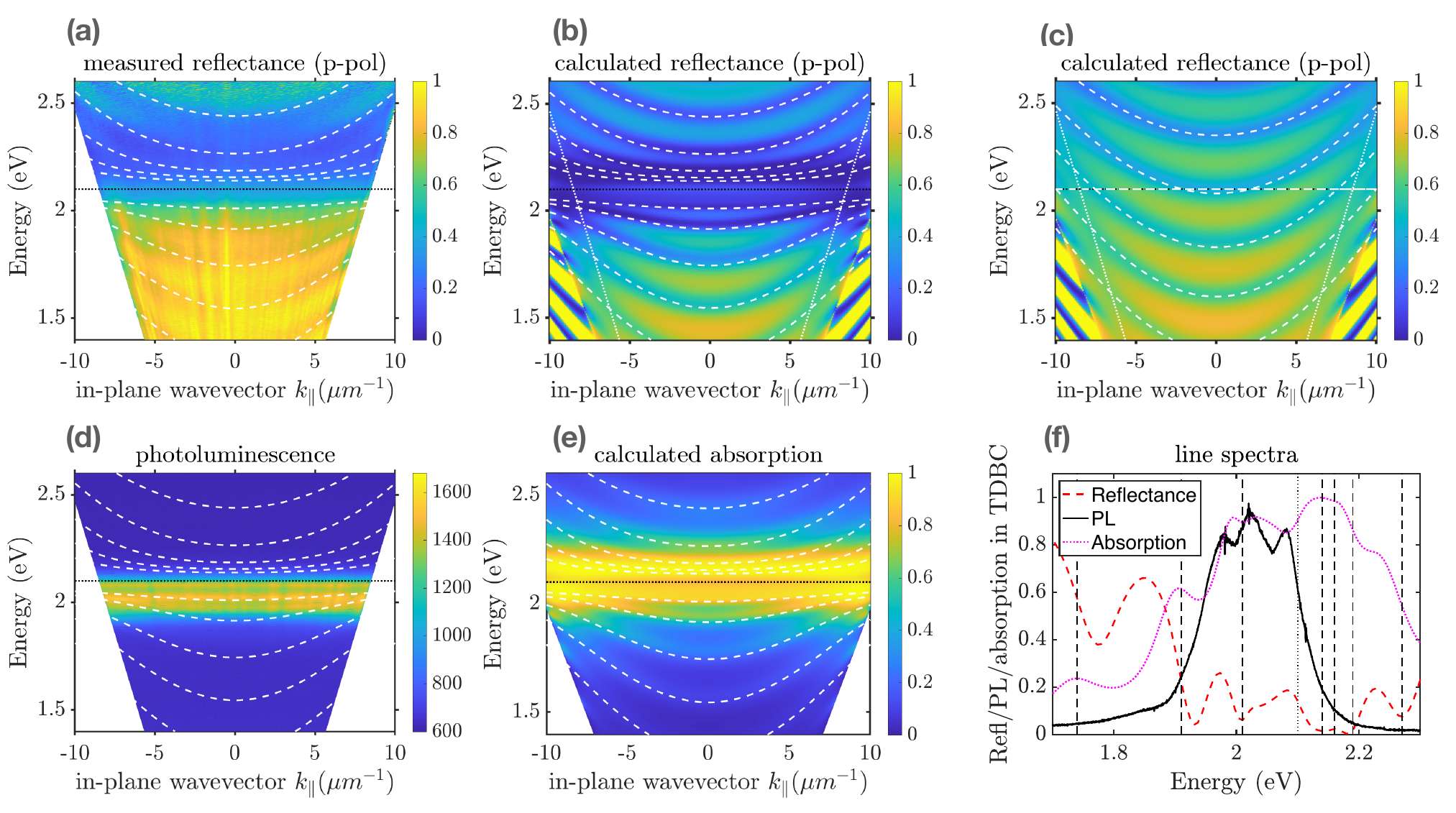}
\caption{{\textbf{Half Cavity (1630 nm)}.
In panel (a) we show an experimentally measured dispersion, based on p-polarised reflectance, indicated as a colour plot.
In panel (b) we show the calculated reflectance, the white dotted lines represent the limit of the collection numerical aperture used in the experiment. Panel (c) is the same as panel (b) except that in the calculation we have set the oscillator strength of the dye to zero.
In panel (d) we show an experimentally measured dispersion plot acquired from photoluminescence data. In panel (e) we show the calculated absorption in the TDBC-doped layer. In panels (a-e) the polariton dispersion is shown, as determined from the coupled oscillator model described in the methods section. The polariton positions are shown as white dashed lines, and the molecular resonance (2.10 eV) by a black dotted line; three cavity modes were used in the model (see panel (c)), and the polymer thickness was $\sim$1630 nm.} In panel (f) the polariton positions (normal incidence) are indicated by black dashed lines.}
\label{fig:SI_half_data_1630}
\end{figure}

\vspace{0.3cm}
Figure \ref{fig:SI_half_data_1630} shows data from a 1630 nm TDBC-doped PVA layer on a gold-coated silicon substrate. Looking first at the reflectance, panels (a) and (b), the presence of modes of the half cavity system are again indicated by regions of low reflectivity (blue). Also shown are the positions of the polaritons as determined by our coupled oscillator model, where we made use of three cavity modes. We see that overall there is a reasonable match between the experimental data, the calculated data, and the coupled oscillator model. The calculated dispersion for a `bare' cavity and the `bare' coupled oscillator model is shown in panel (c).\\

From the modelling we determine that in this case there are five photonic modes to consider, with de-tunings (relative to the molecular resonance of -0.75, -0.50, -0.27, -0.02, and +0.25 eV, having beta values of 1.1, 1.1, 1.1, 1.1, and 1.0 respectively. We determine that in this case the reduced oscillator strength is 0.06, the splitting is $\Omega = 0.35\pm0.02$ eV. 
Panel (d) shows the measured photoluminescence dispersion. As for the 600 nm thick half cavity, things are now slightly more interesting than they were for the open cavity. The PL spectrum is somewhat different from the open cavity case, and there is perhaps some slight mapping of the PL onto the position of the lower polariton modes. Calculated data for the absorption in the TDBC are shown in panel (e). There is again a significant change in the absorption compared to that of the bulk (a single absorption peak), and further there is now an indication of the absorption tracking the polariton modes, at least in some very limited way.

\clearpage

\section{\textbf{Full Cavities}}

\begin{figure}[htb!]
\centering
\includegraphics[width=1.00\linewidth]{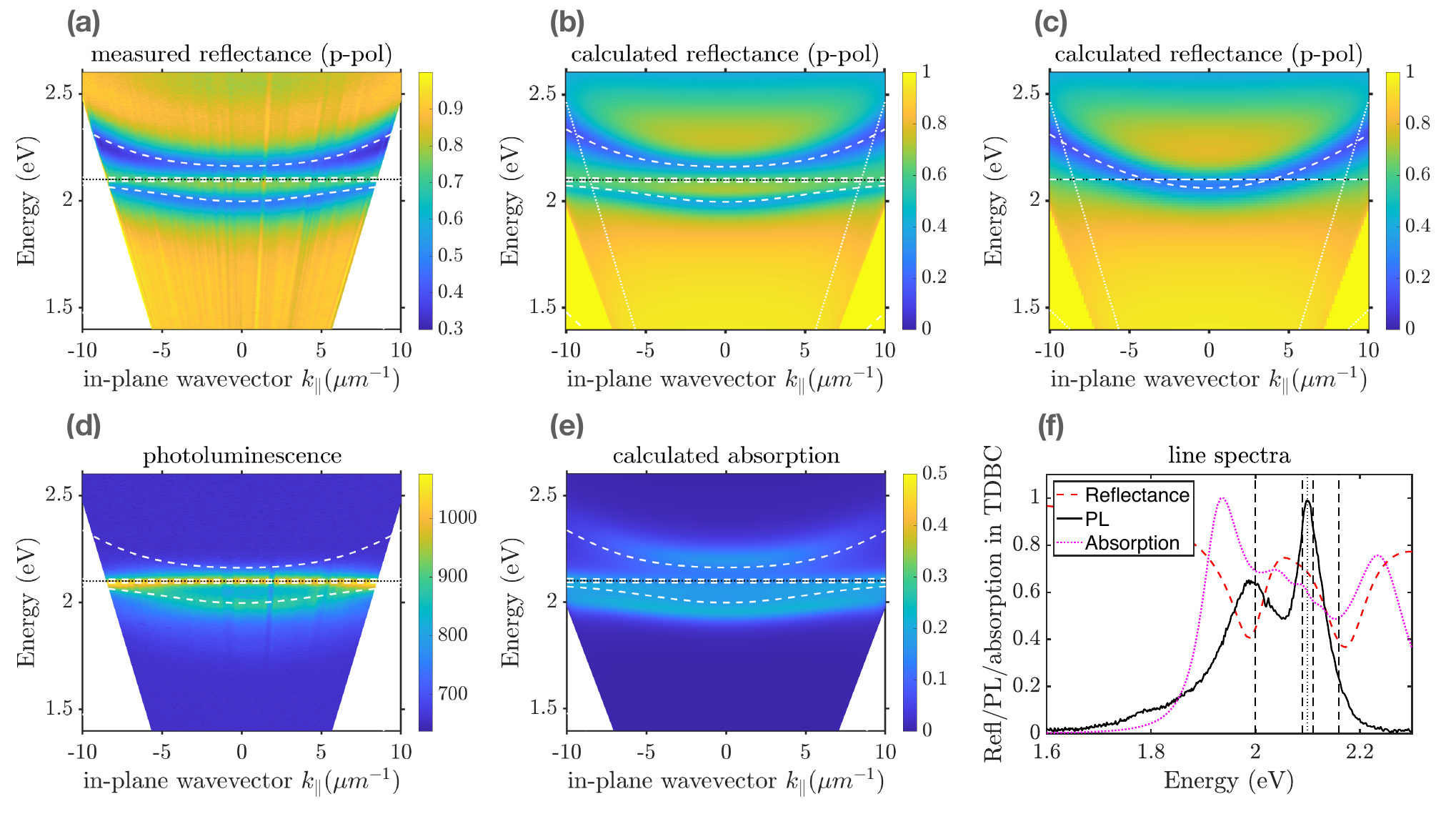}
\caption{\textbf{Full cavity}, cavity thickness 330 nm.
(a) Experimentally measured dispersion plot, based on p-polarised reflectance, indicated as a colour plot. (b) Calculated reflectance, the white dotted lines again represent the limit of the collection optics used in the experiment. (d) Photoluminescence. (e) Calculated absorption in the TDBC. In all plots the polariton dispersion is shown, as determined from the coupled oscillator model described in the methods section. The polariton positions are shown as white dashed lines, and the molecular resonance at 2.10 eV by a black dashed line; only one cavity mode was used in the model. (f) Normal incidence line spectra for the calculated reflectance, the measured photoluminescence and the calculated absorption in the TDBC. The polariton positions (normal incidence) are indicated by black dashed lines.}
\label{fig:half_data_600}
\end{figure}

\vspace{0.3cm}
Again, as for the open and half cavities, in figure S6 
we show the reflectance, panels (a) and (b), the presence of modes of the full cavity system are again indicated by regions of low reflectivity (blue). Also shown are the positions of the polaritons as determined by our coupled oscillator model, where we made use of just one cavity mode. We see that overall there is a reasonable match between the experimental data, the calculated data, and the coupled oscillator model. The calculated dispersion for a `no-resonance' cavity and the `no-resonance' coupled oscillator model is shown in panel (c). In panel (f) we show normal incidence line spectra for the calculated reflectance, the measured photoluminescence and the calculated absorption in the TDBC.
The calculated dispersion for a `no-resonance' cavity and the `no-resonance' coupled oscillator model are shown in panel (c).

\vspace{0.3cm}

From the modelling we determine that in this case there are three photonic modes to consider, with de-tunings (relative to the molecular resonance of -1.04, -0.04, and +0.72 eV, all having beta values of 0.65. The reduced oscillator strength is 0.018, the splitting is $\Omega_R = 0.16\pm0.01$ eV, and $\beta = 0.65$. The width of the `no-resonance' cavity modes in this case is $K = 0.13\pm0.02$ eV
.\\

For the two cavities that follow, the fabrication procedure was slightly different. We made use of a layer-by-layer deposition technique to place 4 monolayers of TDBC molecules (total 8 nm thick) inside these cavities, as follows. A 100 nm thick layer of PMMA was added on top of the first 40 nm gold mirror by spin coating, as to create part of the cavity. Then the TDBC film was deposited on top of this PMMA layer using a layer-by-layer approach~\cite{27}. Briefly, we used a cationic poly(diallyldimethylammonium chloride) (PDAC) solution as the polyelectrolyte binder for anionic TDBC J - aggregate solution. A typical deposition step consists of subsequent dipping the substrate inside a beaker of PDAC solution (20$\%$ by weight in water - diluted 1:1000) and TDBC solution in water (0.01M  diluted 1:10) for 15 minutes each. The substrate was washed with DI water after each immersion and same steps were repeated to deposit multiple layers of PDAC/TDBC. To increase the adhesion we first coat one layer of anionic polystyrene sulfonate (PSS) using the above mentioned process and continue with PDAC - TDBC. Finally the TDBC layer was protected by depositing a layer of PDAC molecules. The superstrate was then prepared by spinning a second layer of PMMA and finally the top mirror was prepared by thermally evaporating a further 40 nm of gold.

\begin{figure}[h!]
\centering
\includegraphics[width=1.00\linewidth]{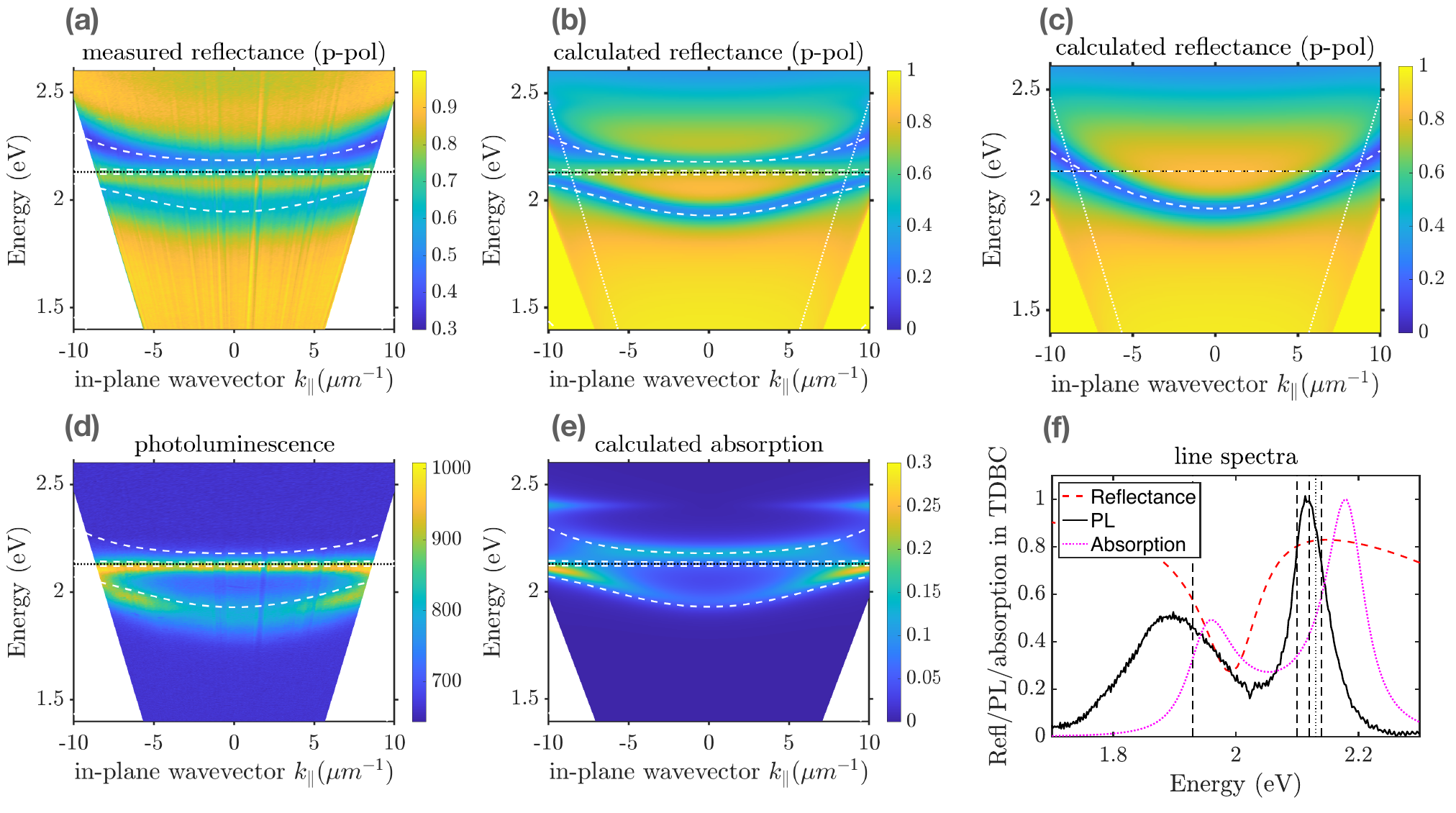}
\caption{{\textbf{Full Cavity (355 nm)}.
In panel (a) we show an experimentally measured dispersion, based on p-polarised reflectance, indicated as a colour plot.
In panel (b) we show the calculated reflectance, the white dotted lines represent the limit of the collection numerical aperture used in the experiment. Panel (c) is the same as panel (b) except that in the calculation we have set the oscillator strength of the dye to zero.
In panel (d) we show an experimentally measured dispersion plot acquired from photoluminescence data. In panel (e) we show the calculated absorption in the TDBC-doped layer. In panels (a-e) the polariton dispersion is shown, as determined from the coupled oscillator model described in the methods section. The polariton positions are shown as white dashed lines, and the molecular resonance (2.13 eV) by a black dotted line; just one cavity mode was used in the model (see panel (c)), and the cavity thickness was $\sim$355 nm.} In panel (f) the polariton positions (normal incidence) are indicated by black dashed lines.}
\label{fig:full_data_355}
\end{figure}

\begin{figure}[h!]
\centering
\includegraphics[width=1.00\linewidth]{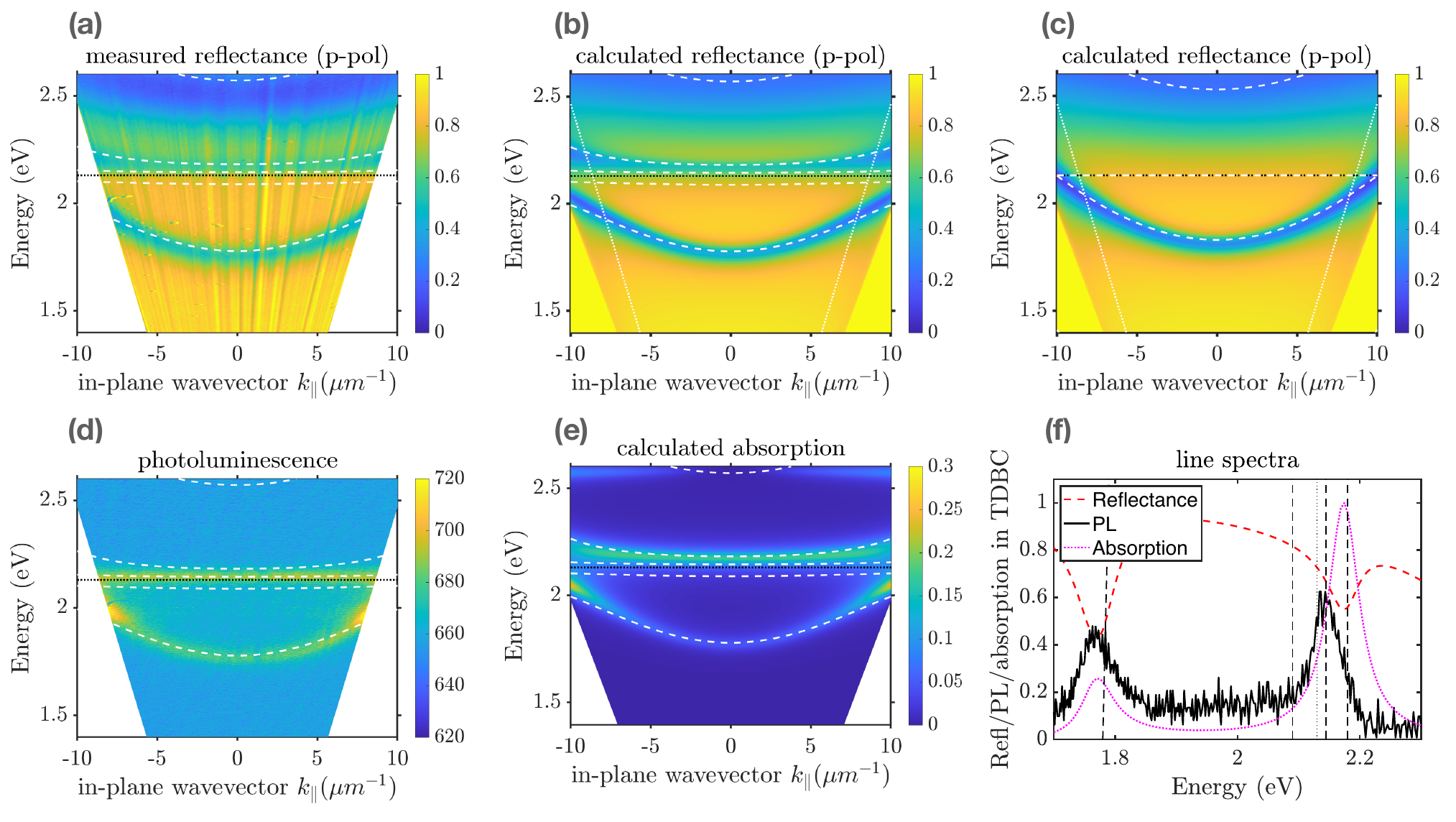}
\caption{{\textbf{Full Cavity (400 nm)}.
In panel (a) we show an experimentally measured dispersion, based on p-polarised reflectance, indicated as a colour plot.
In panel (b) we show the calculated reflectance, the white dotted lines represent the limit of the collection numerical aperture used in the experiment. Panel (c) is the same as panel (b) except that in the calculation we have set the oscillator strength of the dye to zero.
In panel (d) we show an experimentally measured dispersion plot acquired from photoluminescence data. In panel (e) we show the calculated absorption in the TDBC-doped layer. In panels (a-e) the polariton dispersion is shown, as determined from the coupled oscillator model described in the methods section. The polariton positions are shown as white dashed lines, and the molecular resonance (2.13 eV) by a black dotted line; just one cavity mode was used in the model (see panel (c)), and the cavity thickness was $\sim$400 nm.} In panel (f) the polariton positions (normal incidence) are indicated by black dashed lines.}
\label{fig:full_data_400}
\end{figure}

\vspace{0.3cm}

We discuss the results for both the 355 nm and 400 nm cavities at the same time since they are very similar. Looking first at the reflectance, panels (a) and (b), the presence of modes of the full cavity system are again indicated by regions of low reflectivity (blue). Also shown are the positions of the polaritons as determined by our coupled oscillator model, where we made use of one cavity mode (see panel (c)). We see that overall there is a reasonable match between the experimental data, the calculated data, and the coupled oscillator model. The calculated dispersion for a `bare' cavity and the `bare' coupled oscillator model is shown in panel (c).

From the modelling we determine that in both cases there are three photonic modes to consider:

\noindent  for the 355 nm cavity, the de-tunings are -1.13, -0.15, and +1.06 eV, all having beta values of 0.65. The reduced oscillator strength is 0.63, and the splitting is $\Omega_R = 0.20\pm0.02$ eV; for the 400 nm cavity, the de-tunings are -1.23, -0.30, and +0.40 eV, all having beta values of 0.7. The reduced oscillator strength is 1.08, and the splitting is $\Omega_R = 0.27\pm0.02$ eV.

Panel (d) in figures S7 and S8 both show the measured photoluminescence dispersion, the PL spectrum is very clearly different from the open and half cavity cases, the PL nicely mapping onto the position of the lower polariton mode. Calculated data for the absorption in the TDBC are shown in panel (e). There is again a very significant change in the absorption compared to that of the bulk (a single absorption peak), the absorption now tracking the polariton modes.

\clearpage

\section{\textbf{Estimating the Free Spectral Range and the cavity mode linewidth}}

The bare cavity mode linewidth was estimated from the calculated reflectance data with the oscillators turned off, e.g. panel C of figure S2 etc.. The estimate was based on the FWHM at an in-plane wavevector that corresponded to the crossing point with the excitonic resonance.\\

\noindent The Free Spectral Range was estimated from the average of the frequency difference to the next nearest modes, i.e. one mode higher in frequency, one mode lower in frequency.

\clearpage

\section{\textbf{Photoluminescence from different thin TDBC layers}}

\begin{figure}[h!]
\centering
\includegraphics[width=0.9\linewidth]{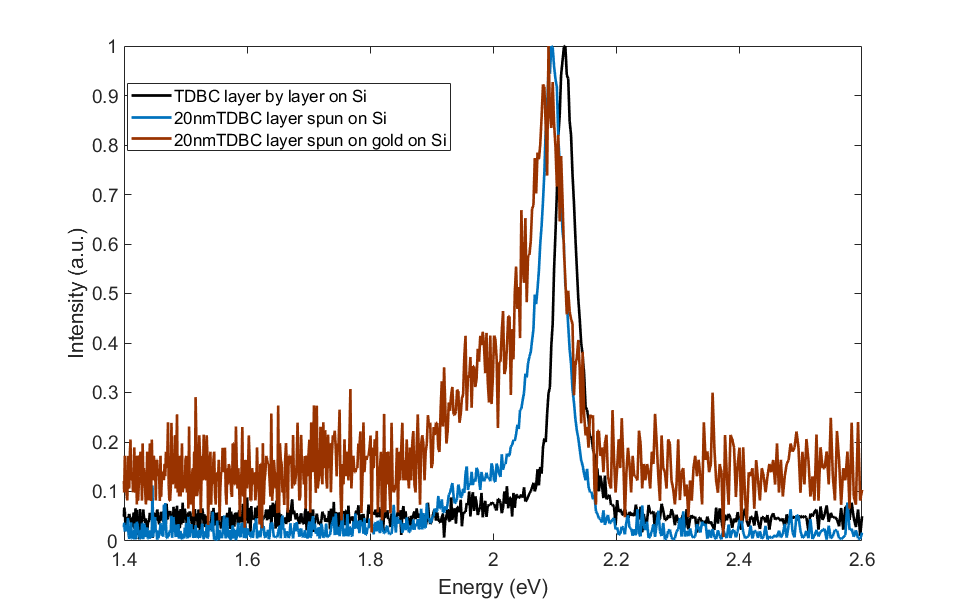}
\caption{\textbf{Photoluminescence from thin TDBC layers}. Three spectra are shown: PL from TDBC produced by the layer-by-layer technique, deposited on Si; PL from TDBC-doped PVA on Si; PL from TDBC-doped PVA on Au coated Si.}
\label{fig:SI_PL_data}
\end{figure}

As a comparison for the PL data we recorded from the different cavities, in figure~\ref{fig:SI_PL_data} we show the PL spectra of thin films of TDBC-doped material.

\vspace{0.3cm}
Note that the PL for TDBC-doped PVA peaks at a lower energy ($\sim$2.10 eV) than the TDBC produced using the layer-by-layer technique ($\sim$2.13 eV). Note also that for the TDBC-doped PVA on Au coated Si, the low energy shoulder of the TDBC PL is enhanced in relative strength. This is consistent with a recent observation based on squarine dyes~\cite{Quenzel_ACSNano_2022_16_4693}. We further note that by using ellipsometry~\cite{Rider_NanoPhot_2024} we determined that the resonance energy for the layer-by-layer TDBC film is at 2.14eV, i.e. to within experimental error it is the same as the 2.13eV of the PL reported here.

\clearpage

\section{\textbf{Literature Values}}

\noindent Data used for the numbered points in figure 3 are as follows:

\noindent 1) Wersall \textit{et al.}~\cite{Wersall_NL_2016_16_551}. The numbers we extracted from this work for inclusion in the present manuscript are:
$\Omega_R=0.23\pm0.03$ eV, FSR = 2.1$\pm0.1$ eV (we assume the nano-prism only supports one relevant plasmonic mode), $K=0.21\pm0.03$ eV, $\Gamma=0.10\pm0.01$ eV. These lead to a value for the finesse of $\mathcal{F} = 10\pm2$, and a value for $2\Omega_R/(\Gamma+K)$ of 1.5 $\pm0.2$.\\


\noindent 2) Thomas \textit{et al.}~\cite{Thomas_JPCL_2021_12_6914}. The numbers we extracted from this work for inclusion in the present manuscript are: 
$\Omega_R=0.59\pm0.03$ eV, FSR = 1.2$\pm0.1$ eV, $K=0.46\pm0.01$ eV (note this is a more recent estimate than the originally published data), $\Gamma=0.36\pm0.01$ eV. These lead to a value for the finesse of $\mathcal{F} = 2.6\pm0.4$, and a value for $2\Omega_R/(\Gamma+K)$ of 1.4 $\pm0.2$.\\

\noindent 3) Vasista \textit{et al.}~\cite{Vasista_NL_2020_20_1766}. The numbers we extracted from this work for inclusion in the present manuscript are: $\Omega_R=0.094\pm0.01$ eV, FSR = 0.10$\pm0.01$ eV, $K=0.03\pm0.005$ eV, $\Gamma=0.11\pm0.01$ eV. These lead to a value for the finesse of $\mathcal{F} = 3.3\pm1.1$, and a value for $2\Omega_R/(\Gamma+K)$ of 1.4 $\pm0.2$.\\

\noindent 4) Thomas \textit{et al.}~\cite{Thomas_Science_2019_363_615}. The numbers we extracted from this work for inclusion in the present manuscript are: $\Omega_R=70\pm5$ cm$^{-1}$, FSR = 420$\pm10$  cm$^{-1}$ $K=30\pm5$ cm$^{-1}$, $\Gamma=30\pm2$ cm$^{-1}$. These lead to a value for the finesse of $\mathcal{F} = 14.0\pm0.4$, and a value for $2\Omega_R/(\Gamma+K)$ of 2.3 $\pm0.1$.\\ 

\noindent 5) Ahn \textit{et al.}~\cite{Ahn_Science_2023_380_6650}. The numbers we extracted from this work for inclusion in the present manuscript are: $\Omega_R=73\pm6$ cm$^{-1}$, FSR = 480$\pm10$  cm$^{-1}$ $\omega_0=2260$ cm$^{-1}$, $K=28\pm6$ cm$^{-1}$, $\Gamma=47\pm1$ cm$^{-1}$. These lead to a value for the finesse of $\mathcal{F} = 11.8\pm0.3$, and a value for $2\Omega_R/(\Gamma+K)$ of 2.6 $\pm0.2$.\\

\section{Theory for Microcavity PL with Finite Finesse}\label{app:PL theory}

The derivation of Eq. (3) for the photoluminescence (PL) spectrum of the  lower polariton is given, starting from a microscopic quantum model. The stationary photoluminescence spectrum $S(\omega)$ of a microcavity is given by the two-time autocorrelation function of the $q$-th cavity mode ~\cite{Herrera_PRA_2017_95_053867}
\begin{equation}\label{eq:noise spectrum}
S_{q}(\omega)=\int_0^{\infty} d \tau\langle\hat{a}_q^{\dagger}(\tau) \hat{a}_q(0)\rangle e^{i \omega \tau},
\end{equation}
where $\hat a_q$ is the annihilation operator of the mode. Using the quantum regression formula, Eq. (\ref{eq:noise spectrum}) can be evaluated analytically in the polariton basis in the approximation that no quantum jumps occur, to give~\cite{Herrera_PRA_2017_95_053867} 
\begin{equation}\label{eq:PL eigenstates}
S_{q}(\omega)=\sum_{i j} \rho_j\left|\left\langle\epsilon_i|\hat{a}_q| \epsilon_j\right\rangle\right|^2 \frac{\left(\Gamma_j / 2\right)}{\left(\omega-\omega_{j i}\right)^2+\left(\Gamma_j / 2\right)^2},
\end{equation}
where $\rho_j$ is the stationary population of the coupled system  eigenstate $|\epsilon_j\rangle$, $\Gamma_j$ is the decay rate of $|\epsilon_j\rangle$, and $\omega_{ji}=\epsilon_j-\epsilon_i$ is the  frequency of the transition $|\epsilon_j\rangle\rightarrow|\epsilon_i\rangle$, with $\epsilon_j>\epsilon_i$ being the state energies. Considering only the the ground and first excitation manifolds, which determine the linear response of the system, Eq. (\ref{eq:PL eigenstates}) can be written as
\begin{equation}\label{eq:PL single excitation}
 S_{q}(\omega)=\pi \sum_{j} \rho_j\left|\langle G|\hat{a}_q| \epsilon_j\rangle\right|^2 L_j(\omega)
\end{equation}
where $\ket{G}=\ket{g_1,g_2,\ldots,g_N}\ket{\{0_q\}}$ is the global ground state with all dipoles in a local ground state $\ket{g_n}$ and all cavity modes in the vacuum state. $L_j(\omega)$ is a normalized Lorentzian lineshape function centered at $\omega_j$ with bandwidth $\Gamma_j$ (FWHM), associated with the transition $G\rightarrow{\epsilon_j}$.

As shown by Herrera and Spano~\cite{Herrera_PRA_2017_95_053867}, solving for the eigenstates and energies of all the polaritonic and dark states of the coupled light-matter system in the single excitation manifold, enables the direct evaluation of Eq. (\ref{eq:PL single excitation}), up to an undetermined excited population parameter $\rho_j$. Instead of assuming an arbitrary state-independent value $\rho_j\sim 1/N$~\cite{Herrera_PRA_2017_95_053867}, we derive an  analytical expression for $\rho_j$ following the observation that ultraviolet pumping of material dipoles followed by ultrafast energy relaxation down to the lowest singlet ($S_1$) excited state is equivalent to the direct incoherent excitation of local molecular dipoles with jump operator $\hat L_j=\sqrt{W}\hat \sigma_{+n}$, where $\hat \sigma_{+n}$ creates an excitation in the $n$-th molecular dipole and $W$ is the effective incoherent pumping rate. The proposed equivalence can be justified using standard adiabatic elimination techniques~\cite{Zhu_2015}.

The quantum master equation that describes incoherent dipole pumping and cavity decay can be written in Lindblad form as 
\begin{equation}\label{eq:lindblad equation}
\frac{d \hat{\rho}}{d t}=-i[\hat{H}, \hat{\rho}]+\sum_n W_n\left(\hat{\sigma}_{+n}\hat{\rho}\hat{\sigma}_{-n}-\frac{1}{2}\left\{\hat{\sigma}_{-n} \hat{\sigma}_{+n}, \hat\rho\right\}\right) 
+\sum_q \kappa_q\left(\hat{a}_q \hat\rho \hat{a}_q^{\dagger}-\frac{1}{2}\left\{\hat a_q^{\dagger} \hat a_{q} \hat{\rho}\right\}\right),
\end{equation}
where $\hat H$ is the coupled light-matter Hamiltonian representing $N$ two-level dipoles in a multimode microcavity, $W_n$ is the local excitation rate of the $n$-th dipole and $\kappa_q$ is the decay rate of the $q$-th cavity mode. $\{\hat A,\hat B\}$ denotes an anticommutator. For long times $t\gg 1/W_n$, we expect the incoherently driven system to evolve into mixed state $\hat\rho=\sum_j\rho_j\ket{\epsilon_j}\bra{\epsilon_j}$, with state population $0<\rho_j<1$. Equations of motion for the population of coupled eigenstates $\rho_j^{(1)}$ in the first-excitation manifold are obtained directly from Eq. (\ref{eq:lindblad equation}) in the mixed-state ansatz $\hat\rho=(1-\chi_W)\ket{G}\bra{G}+\sum_j\rho_j^{(1)}\ket{\epsilon_j}\bra{\epsilon_j}$, giving
\begin{equation}\label{eq:population equation}
\frac{d}{d t}\rho_j^{(1)}  =W \sum_n\left|\langle\epsilon_j|\hat \sigma_n^{+}| G\rangle\right|^2\left(1-\chi_W\right) - W \sum_n\langle\epsilon_j|\sigma_n^{-} \sigma_n^{+}| \epsilon_j\rangle \rho_j^{(1)} -\kappa \sum_q\langle\epsilon_j|\hat a_q^{+} \hat a_q| \epsilon_j\rangle \rho_j^{(1)},
\end{equation}
where homogeneous pumping and mode-independent bandwidth conditions were assumed ($W_n=W$ and $\kappa_q=\kappa$). The ground depletion parameter $\chi_W=\sum_j\rho_j^{(1)}$ is introduced for completeness, but can be neglected for the weak pumping conditions $W\ll \kappa$ relevant to this work. Although the second-moment $\langle\epsilon_j|\sigma_n^{-} \sigma_n^{+}| \epsilon_j\rangle$ virtually couples the state with the second-excitation manifold, it is kept in the Eq. (\ref{eq:population equation}) because it does not involve double excitation of the same dipole. 

The steady state solution of Eq. (\ref{eq:population equation}) with $\chi_W\rightarrow 0$ is 
\begin{equation}
\rho_{jss}^{(1)}= \frac{W \sum_n\left|\langle\epsilon_j|\hat \sigma_n^{+}| G\rangle\right|^2}{W \langle\epsilon_j|\sum_n\sigma_n^{-} \sigma_n^{+}| \epsilon_j\rangle  +\kappa \langle\epsilon_j|\sum_q\hat a_q^{+} \hat a_q| \epsilon_j\rangle},    
\end{equation}
which can be evaluated explicitly within the single-excitation wavefunction ansatz
\begin{equation}
|\epsilon_j\rangle=\sum_{n} c_{i \epsilon_j}|g_1,\ldots,e_n,\ldots g_N\rangle  |\{0_q\}\rangle +\sum_q c_{q \epsilon_j}|G\rangle|1_q\rangle,
\end{equation}    
to give
\begin{equation}\label{eq:rho ss}
    \rho_{jss}^{(1)}=\frac{W X_j^T}{W(N-1)X_j^T+(NW+\kappa)C^T_j},
\end{equation}
where $X_j^T=\sum_n|c_{n\epsilon_j}|^2$ is the total exciton fraction of the $j$-th coupled eigenstate $\ket{\epsilon_j}$ and $C_j^T=\sum_q |c_{q\epsilon_j}|^2$ is the total photon fraction. Combining Eq. (\ref{eq:rho ss}) with (\ref{eq:PL eigenstates}), the analytical expression for the PL spectrum in the limit $N\rightarrow \infty$, with $NW$ finite, becomes
\begin{equation}\label{eq:PL final}
S_{q}(\omega)= \pi W \,\sum_j \frac{ X_j^T\,C_{jq}}{NW X_j^T+(NW+\kappa)C_j^T}\,L_j(\omega),
\end{equation}
which is Eq. (3) in the main text, were the cavity decay rate is expressed in terms of bandwidth $K=2\kappa$ (FWHM). In the limit where photon loss dominates over the excitation rate, i.e., $\kappa\gg NW$ and the free spectral range is infinite, $C_j^T=C_j^q$, the cavity PL spectrum  reduces to $S_T(\omega)\approx\pi W\sum_jX_j^T L_j(\omega)$, i.e., is fully determined by the exciton content at particular system eigenfrequencies $\omega_j$. 

\begin{figure}
    \centering
    \includegraphics[width=\textwidth]{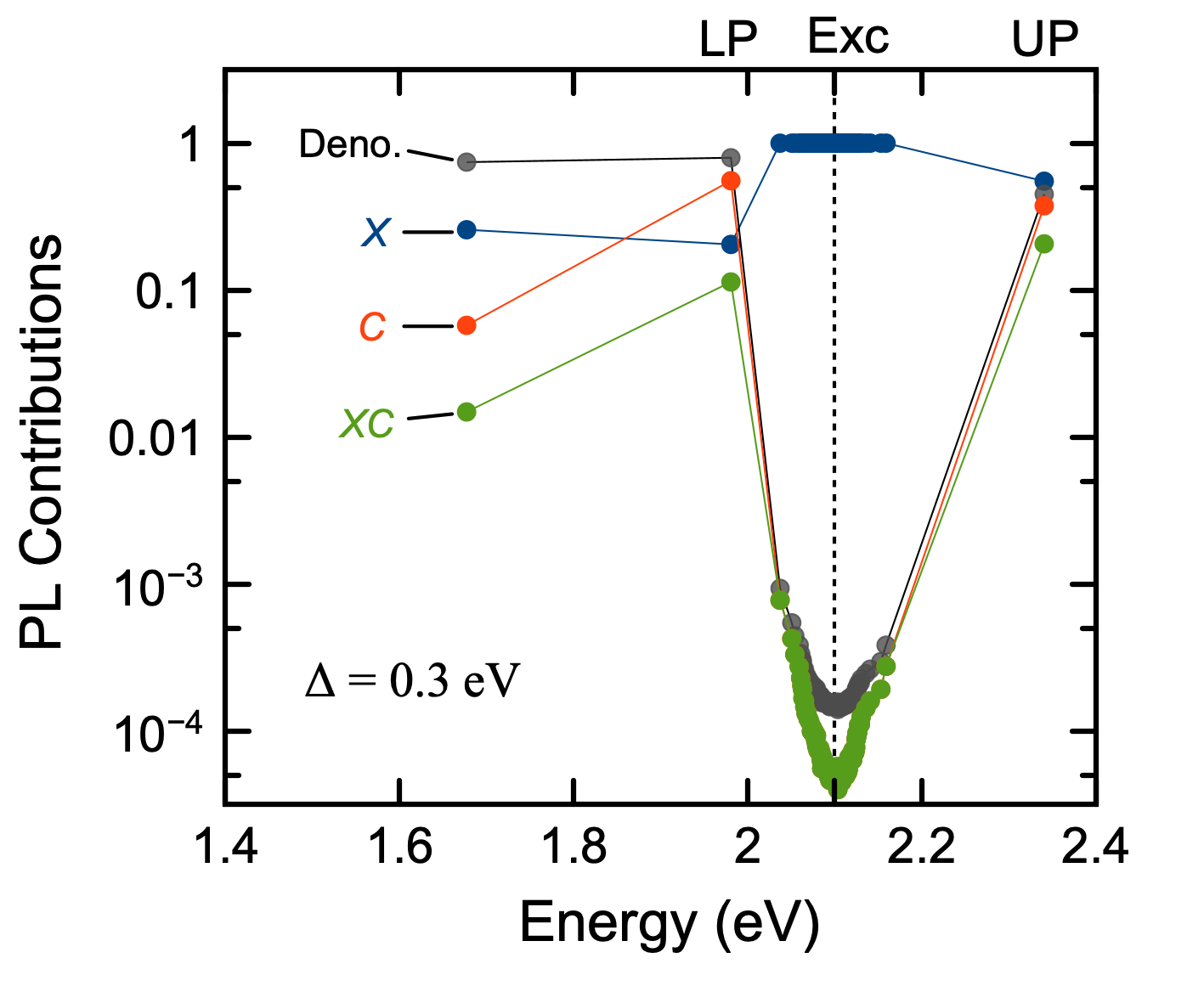}
    \caption{\textbf{Spectrum of contributions to the expression for PL}, including total exciton content $X$ [$X_j^T$ in Eq. (\ref{eq:PL final})],  total photon content $C$ ($C_j^T$), light-matter product $XC$ ($X_j^T\,C_{jq}$), and denominator $NW X_j^T+(NW+\kappa)C_j^T$, for a two-mode cavity with (low) finesse $\Delta =0.3$ eV that couples to a Gaussian distribution of exciton transitions (Exc) centered at 2.1 eV ($\sigma$ =0.05 eV, $\kappa=K/2= 0.1$ eV). Contributions are averaged over 650 disorder configurations.}
    \label{fig:SI_PL_contributions}
\end{figure}

Figure \ref{fig:SI_PL_contributions} shows a spectrum of the terms that contribute to the PL signal in Eq. (\ref{eq:PL final}): total exciton content $X_j^T$,  total photon content $C_j^T$, the product $X_j^T\,C_{jq}$, and the denominator $NW X_j^T+(NW+\kappa)C_j^T$, for each light-matter eigenstate in a model system composed of an ensemble of $N=150$ excitonic dipoles with Gaussian disorder centered at $\omega_e=2.1$ eV  ($\sigma=0.05$ eV) interacting with a two-mode cavity with mode frequencies $\omega_{q-1}=1.8$ eV and $\omega_{q}=2.1$ eV, both having equal bandwidth $\kappa =0.1$ eV. Each molecular dipole couples to both cavity modes with single-molecule Rabi frequency $g = 0.2/\sqrt{N}$ eV. We use $NW = 0.0001$ and average over 650 disorder configurations, consistent with the results in Fig. 7 of the main text.

\section{Sample preparation}

\noindent {\textbf{Solution preparation:}}
Poly Vinyl Acetate (PVA, molar weight 450 000) was used as a host matrix for TDBC.
PVA was dissolved in water. TDBC was then dissolved in the PVA-water solution.
TDBC/PVA films were deposited on a silicon wafer by spin-coating. This produced film thicknesses over the range 300-1700 nm.

For the open-cavity structures, a TDBC/PVA solution was spun on top of a silicon substrate. For the half-cavity a thin gold layer of 30 nm was deposited on top of silicon substrate by thermal evaporation. Later, TDBC/PVA was spun on top of gold thin film. For the full-cavity, the lower mirror was prepared by thermally evaporating 40 nm of gold onto a silicon substrate. Then, following deposition of the TDBC/PVA layer, a second 40 nm gold mirror was deposited, again by thermal evaporation. 

\section{Reflectance Measurements}

Whilst substrate-only spectra for the reflectance should provide a means to normalise the reflectance data, in our samples a non-trivial level of scattering meant that the normalisation process was not as effective as hoped. Accordingly the reflectance spectra were scaled to provide a better match with calculated data.

\section{Coupled Oscillator Models}

We made use of a standard coupled oscillator model to help us interpret our data~\cite{Richter_APL_2015_107_231104}. The number of cavity resonances included in each model depended on the details of the sample (thickness etc.). For a molecular resonance at a frequency (angular) of  $\omega_{\textrm{mol}}$ interacting with several cavity resonances (whose energy depend on in-plane wavevector, i.e. $\omega_{\textrm{cav}}(k_{\parallel})$), the coupled oscillator model allows the dispersion of the new hybrid (polariton) modes to be determined. Since all of our samples supported more than one cavity modes we made use of a 2N multi-photonic-mode coupled oscillator model~\cite{Richter_APL_2015_107_231104,Balasubrahmaniyam_PRB_2021_103_L241407} to find the polariton frequencies by solving for those frequencies $\omega$ for which the determinant of the coupling matrix is zero, i.e. $|\mathbf{M}|=0$. Although this choice (2N) is not universal, many authors preferring an N+1 coupled oscillator model, we find it to be more universally applicable. As an example, the coupling matrix for one molecular resonance interacting with two cavity (photonic) modes is given by,

\begin{align}
    \mathbf{M} = \begin{pmatrix}\omega_{\textrm{mol}}- \omega & \Omega_R/2 &0&0\\
    \Omega_R/2 & \omega_{1}(k_{\parallel})-\omega&0&0\\
    0&0&\omega_{\textrm{mol}}- \omega & \Omega_R/2
    \\
    0&0&\Omega_R/2 & \omega_{2}(k_{\parallel})-\omega\end{pmatrix}
\label{CO_matrix}
\end{align}

\noindent where $\Omega_R$ is the extent of the interaction (assumed to be the same for all  modes), and is equal to the frequency splitting (anti-crossing) at the value of $k_{\parallel}$ where the uncoupled resonances cross. For the $n^{th}$ cavity resonance we described the $k_{\parallel}$ dependence, $\omega_n (k_{\parallel})$, as,

\begin{equation}
\omega_n = \frac{\omega_{n\,{k_0 = 0}}}{k_0} \sqrt{{k_0}^2 + \beta_n\, {k_{\parallel}}^2/{\varepsilon_b}},
\end{equation}
\label{eq:cavity_k}

\noindent where $\omega_{n\,{k_0 = 0}}$ is the angular frequency of the n$^{th}$ resonance at $k_{\parallel} = 0$. The factor $\beta_n$ was used to match the dispersion of the bare cavity modes with the calculated bare reflectance (needed because the coupled oscillator model ignores the angle-dependant phase change on reflection from an interface). We chose to work with the simplest model that we could, and therefore assumed no damping of the modes. The values of the $\omega_{n\,{k_0 = 0}}$, of $\Omega_R$ and of $\beta_n$ were determined by simultaneously trying to match the eigenvalues to the calculated and measured reflection data.
The values for the Rabi splitting we report here are based on the parameters used in the coupled oscillator model that best match our data.

\section{Transfer Matrix modelling}

We made use of a standard multi-layer optics approach to calculations of the reflectance and absorption~\cite{Smith_JMO_2008_55_2929,NandH,WJTan_PhD}, and assumed all media were isotropic. The reflectance was calculated by taking the square of the modulus of the amplitude reflection coefficient. For the permittivities, we used the following:

\vspace{0.2cm}
\noindent \textbf{\textit{Silicon substrate}}: we took the permittivity to be 16.0 + 3.0i.

\vspace{0.2cm}
\noindent \textbf{\textit{Gold films}}: we used a Drude-Lorentz model for the frequency/energy dependant permittivity of gold, $\varepsilon_{\rm{Au}}(\omega)$, given by~\cite{Rakic_AO_1998_37_5271},

\begin{equation}
\varepsilon_{\rm{Au}}(\omega) = \frac{f_d\,\omega_p^2}{\omega^2+i\gamma_d\omega} + \sum_j{\frac{f_j\omega_p^2}{\omega_j^2-\omega^2-i\gamma_j\omega}},
\end{equation}
\label{eq:eps_Au}

\noindent where the plasma frequency $\omega_p$ was 9.03 eV,  the Drude damping rate $\gamma_d$ was 0.053 eV, and the strength, $f_d$ was 0.76. The Lorentz oscillator parameters we used are,

\begin{table}[htb!]
\centering
\begin{tabular}{| c| c| c| c |}
\hline
$j$   & $f_j$ &$\omega_j (eV)$ & $\gamma_j(eV)$ \\
\hline
1 &$0.024$ & $0.451$ & $0.241$ \\
\hline
2 &$0.010$ & $0.830$ & $0.345$ \\
\hline
3 &$0.071$ & $2.969$ & $0.870$ \\
\hline
4 &$0.601$ & $4.304$ & $2.494$ \\
\hline
5 &$4.385$ & $13.32$ & $2.214$ \\
\hline
\end{tabular}
\caption{Lorentz oscillator model parameters for gold films, as determined from ellipsometry}
\label{tab:eps_Au_parameters}
\end{table}

\vspace{0.2cm}
\noindent \textbf{\textit{TDBC J-aggregates}}: we used a Lorentz model, with the permittivity $\varepsilon_{\rm{TDBC}}(E)$ given by,

\begin{equation}
\label{eq:eps_TDBC}
\varepsilon_{\rm{TDBC}}(\omega) = \varepsilon_b + \frac{f\,\omega^2_0}{\omega^2_0-\omega^2-i\gamma\omega},
\end{equation}

\section{Technical Details re: TOC figure}

The two spectra show the calculated transmission spectra for two of the cavities we investigated, one of low finesse ($\sim 2$) and one of higher finesse ($\sim 8$). Note that the spectrum for the lower finesse cavity shows modes that overlap.

The two curves shown in the TOC figure are the zero-oscillator normal incidence transmittance (calculated) for the following two cavity structures.

We show the calculated transmittance (empty cavity, i.e. zero oscillator strength) for two of our samples, the 600 nm half cavity (dotted line) and the 330 nm full cavity (full line).

Half cavity (600 nm), dotted red line.

Full cavity (330 nm), full blue line.

\bibliography{finesse}